\documentclass{aa}
\usepackage{times,epsfig,amssymb,amsmath}
\usepackage{natbib}
\usepackage{hyperref}
\usepackage{xcolor}

\newcommand{\chandra}{{\it Chandra}}
\newcommand{\xmm}{{\it XMM-Newton}}

\newcommand{\for}{{\it Forward Method}}
\newcommand{\back}{{\it Backward Method}}

\title{The relation between mass and concentration \\
in X-ray galaxy clusters at high redshift}
\titlerunning{Concentration-mass relation in X-ray galaxy clusters at high redshift}

\author{S. Amodeo\inst{1}\thanks{\emph{Present address:} 
    GEPI, Paris Observatory, 77 av. Denfert-Rochereau, 75014 Paris, France}, 
    S. Ettori\inst{2,3}, 
    R. Capasso\inst{1,4,5}, 
    M. Sereno\inst{2, 1}
} 
\authorrunning{S. Amodeo et al.}

\institute{
 Dipartimento di Fisica e Astronomia, Universit\`a di Bologna, viale Berti Pichat 6/2, 40127 Bologna, Italy
 \and INAF, Osservatorio Astronomico di Bologna, via Ranzani 1, I-40127 Bologna, Italy
 \and INFN, Sezione di Bologna, viale Berti Pichat 6/2, I-40127 Bologna, Italy
 \and Faculty of Physics, Ludwig-Maximilians-Universitaet, Scheinerstr. 1, 81679 Muenchen, Germany
 \and Excellence Cluster Universe, Boltzmannstr. 2, 85748 Garching, Germany 
}

\offprints{S. Amodeo}

\mail{stefania.amodeo@obspm.fr}

\date{}

\begin{document}

\abstract
{Galaxy clusters are the most recent, gravitationally-bound products of the hierarchical mass accretion over cosmological scales.
How the mass is concentrated is predicted to correlate with the total mass in the cluster's halo, with systems at higher mass being less
concentrated at given redshift and for any given mass, systems with lower concentration are found at higher redshifts.}
{Through a spatial and spectral X-ray analysis, we reconstruct the total mass profile of 47 galaxy clusters observed with \chandra\ 
in the redshift range $0.4<z<1.2$, 
selected to have no major mergers, to investigate the relation between 
the mass and the dark matter concentration, and the evolution of this relation with redshift. 
The sample in exam is the largest one investigated so far at $z>0.4$, and is well suited to provide 
the first constraint on the concentration--mass relation at $z>0.7$ from X-ray analysis.
}
{Under the assumptions that the distribution of the X-ray emitting gas is spherically symmetric and in the hydrostatic equilibrium with 
the underlined gravitational potential, we combine the deprojected gas density and spectral temperature profiles through the hydrostatic 
equilibrium equation to recover the parameters that describe a Navarro-Frenk-White total mass distribution. 
The comparison with results from weak lensing analysis reveals a very good agreement both for masses and concentrations. 
Uncertainties are however too large to make any robust conclusion on the hydrostatic bias of these systems.}
{ The distribution of concentrations is well approximated by a lognormal function in all the mass and redshift ranges investigated. 
The relation is well described by the form $c \propto M^B (1+z)^C$, with $B=-0.50 \pm 0.20$, $C=0.12 \pm 0.61$ (at 68.3\% confidence). 
This relation is slightly steeper than the one predicted by numerical  simulations ($B\sim-0.1$) and does not show any evident redshift evolution. 
We obtain the first constraints on the properties of the concentration--mass relation at $z > 0.7$ from X-ray data,
showing a reasonable good agreement with recent numerical predictions.}
{}


\maketitle


\section{Introduction}
Within the standard cosmological model, structure formation takes place from the gravitational collapse of small perturbations in a 
quasi-homogeneus Universe dominated by cold dark matter (CDM). The collapse proceeds from smaller to larger scales giving rise to a 
hierarchical clustering of cosmic structures. In this framework, galaxy clusters, being the largest nearly virialized collapsed objects in the 
observable Universe, are also the last to form. Therefore, they are fundamental tools for understanding the formation and evolution of 
cosmic structures.

Numerical N-body simulations predict that dark matter halos have a universal density profile characterised by two parameters: 
the scale radius $r_\text{s}$, defined as the radius at which the logarithmic density slope is $-2$, and the concentration $c$, defined as the ratio
between $R_{200}$\footnote{$R_{200}$ is the radius within which the cluster density is $200$ times the critical density of the Universe at
the cluster's redshift.} and $r_\text{s}$ \citep[hereafter NFW]{nfw97}. Because of the hierarchical nature of structure formation
(low-mass objects form earlier than high-mass ones)
and the fact that collapsed objects retain information on the background density at the time of their formation (the background average matter
density was higher in the past), concentration and mass are related so that systems with higher masses are less concentrated and,
at a given mass, lower concentrations are expected at higher redshifts \citep[e.g.][]{munoz+11}. 
Moreover, the properties of the background Universe depend on the set 
of cosmological parameters adopted: models with lower matter density and lower normalisation of the linear power spectrum result in a later 
assembly redshift, so less concentrated halos are expected at a given mass. Therefore, the $c(M, z)$ relation contains a wealth of cosmological
information. Several works have been performed to characterise this relation, both numerically and observationally, but there are tensions 
between them. Numerical simulations by \citet{dolag+04}, \citet{duffy+08}, \citet{bhatt+13}, \citet{deboni+13}, \citet{ludlow+14}, 
\citet{dm14} indicate that concentration and mass are anti-correlated for all the mass ranges and redshifts investigated, 
with a mass dependence that is slightly reduced at larger redshift. 
Observations of galaxy clusters at low redshift confirm the expected anti-correlation between $c$ and $M$ 
but they generally find a steeper slope and a higher normalisation compared to the theoretical relation \citep{buote+07,sa07,ettori+10,merten+15}. 
Whether this discrepancy is due to observational selection biases \citep[e.g.][]{meneghetti+14,sereno+15}
or to the lack of some fundamental physics in numerical models is still an open question. Both simulations \citep[e.g.][]{deboni+13} and
observations \citep{ettori+10} agree on the influence of the dynamical state of a cluster on its concentration: more relaxed systems are 
more concentrated, at a fixed mass. A different trend emerge from simulations by \citet{prada+12} and \cite{klypin+14}. 
They predict that at high redshifts the $c(M)$ relation has a plateau and an upturn, at the typical masses of galaxy clusters. 
However, as shown in \citet[][see also \citealt{correa+15}]{ludlow+12}, this plateau/upturn disappears when only the more relaxed halos are considered.
Properties of observed mass-concentration relations are strongly sample-dependent \citep{sereno+15}. The predicted slope in signal-selected samples can be much steeper than that of the underlying population characterising dark matter-only clusters. Over-concentrated clusters can be preferentially included and this effect is more prominent at the low mass end. \citet{sereno+15} found this trend both in the X-ray selected samples CLASH \citep[The Cluster Lensing And Supernova survey with Hubble,][]{pos+al12} and LOCUSS \citep[Local Cluster Substructure Survey,][]{oka+al10} and in the lensing selected sample SGAS \citep[Sloan Giant Arcs Survey,][]{hen+al08}. Statistical and selection biases in observed relations are then to be carefully considered when comparing with predictions of the $\Lambda$CDM model \citep{meneghetti+14}.
Among the methods used to characterise the $c(M)$ relation, X-ray observations are found to be rather
successful since galaxy clusters have a well resolved extended emission with a total luminosity that is proportional to the square of the 
gas density. 

In this work, we perform spatial and spectral analysis for a sample of $47$ galaxy clusters observed with \chandra\
in the redshift range $0.4<z<1.2$, 
selected to have no major mergers,
with the aim to (1) reconstruct their total mass profile by
assuming a spherical symmetry for the intracluster medium (ICM) distribution and hydrostatic equilibrium between the ICM and the gravitational
potential of each cluster, (2) investigate the relation between their mass and concentration and its evolution with redshift. 
Note that we consider here the largest sample investigated so far at $z>0.4$, with the purpose also to probe the $c(M)$ relation at $z>0.7$ 
for the first time using X-ray data.

The paper is organised as follows: in Section~\ref{sec:dataset}, we present the sample of \chandra\ observations selected for the analysis; 
in Section~\ref{sec:analysis} and~\ref{sec:methods}, we describe the data analysis and the method used to reconstruct the clusters mass profiles, 
respectively; in Section~\ref{sec:cm}, we investigate our $c(M,z)$ relation and its redshift evolution.
We discuss the properties of the sample and its representativeness in Section~\ref{sec:properties} 
and we draw our conclusions in Section~\ref{sec:summary}. 
We assume a flat $\Lambda$CDM cosmology with $\Omega_m = 0.3$, $\Omega_\Lambda = 0.7$, $H_0 = 70$ km s$^{-1}$ Mpc$^{-1}$ and 
$h(z) = \sqrt{\Omega_m (1+z)^3 + \Omega_\Lambda}$. 
All quoted errors are $68.3\%$ ($1\sigma$) confidence level, unless otherwise stated.

\section{The dataset}
\label{sec:dataset}
We have retrieved from the \chandra\ public archive all observations of galaxy clusters with redshift $z \geq 0.4$ available at 2nd March 2014.
We have excluded the ones with exposure time shorter than $20$ ks in order to have sufficient X-ray counts statistic, in particular for the 
spectral analysis, and those that to a visual inspection showed evidence of dynamic activity (e.g. presence of major substructures).
This restriction minimizes the systematic scatter in the mass estimate,
since the higher the degree of regular morphology in the X-ray image, the more the cluster is expected to be dynamical relaxed and more robust
is the assumption of the hydrostatic equilibrium of the ICM in the cluster potential well \citep[e.g.][]{rasia+06, poole+06, mahdavi+13, nelson+14}.   
Another selection criterion is related to the choice of adopting a NFW as functional form of the cluster gravitational
profile, which has two free parameters (the scale radius $r_\text{s}$ and the concentration $c$).
Considering that our procedure to reconstruct the mass profile requires independent spectral measurements of the gas temperatures 
(see Section~\ref{sec:methods}), we need a number of independent radial bins larger than the number of mass modelling parameters (=2).
Therefore, we have used only the targets for which we could measure the temperature in at least three independent bins.
The final sample is then composed by $47$ galaxy clusters spanning a redshift range $0.4<z<1.2$, as listed in Table~\ref{tab:sample}.

\begin{table*}
\caption{Sample of the galaxy clusters analysed in this work.}
\vspace*{-0.2cm}
\label{tab:sample}
\begin{center}
\begin{tabular} {c c c c c c c c c c c}
\hline\hline 
Cluster & z & Detector & Exposure [ks] & RA [J2000] & DEC [J2000] & tot cts & $R_\text{out}^\text{spat}$ [kpc] & $n_\text{bin} S_\text{b}$ & 
$R_\text{out}^\text{spec}$ [kpc] & $n_\text{bin} T$ \\
\hline
  MACSJ0159.8-0849 & $0.405$ & ACIS-I &  $29.1$ & $01\ 59\ 49.50$ & $-08\ 49\ 59.3$ & $20250$ & $1130$ & $56$ & $786$ & $13$\\
  MACSJ2228.5+2037 & $0.412$ & ACIS-I &  $16.5$ & $22\ 28\ 32.41$ & $+20\ 37\ 30.5$ & $9234$ & $1511$ & $27$ & $680$ & $5$ \\
   MS1621.5+2640 & $0.426$ & ACIS-I &  $27.5$ & $16\ 23\ 35.40$ & $+26\ 34\ 11.2$ & $9277$  & $1109$ & $20$ & $856$ & $5$ \\
  MACSJ1206.2-0848 & $0.440$ & ACIS-I &  $21.1$ & $12\ 06\ 12.38$ & $-08\ 48\ 07.4$ & $10559$ &  $1131$ & $29$ & $516$ & $5$ \\
  MACSJ2243.3-0935 & $0.447$ & ACIS-I &  $18.5$ & $22\ 43\ 21.57$ & $-09\ 35\ 42.4$ & $9432$  &  $1305$ & $31$ & $537$ & $5$ \\
  MACSJ0329.7-0211 & $0.450$ & ACIS-I &  $28.4$ & $03\ 29\ 41.40$ & $-02\ 11\ 44.4$ & $12870$ &  $950$  & $34$ & $660$ & $8$ \\
  RXJ1347.5-1145 & $0.451$ & ACIS-I &  $29.2$ & $13\ 47\ 30.87$ & $-11\ 45\ 09.9$ & $29013$ &  $1266$ & $66$ & $829$ & $10$ \\ 
      V1701+6414 & $0.453$ & ACIS-I &  $31.1$ & $17\ 01\ 23.41$ & $+64\ 14\ 11.5$ & $9841$ & $892$ & $15$ & $633$ & $6$ \\
  MACSJ1621.6+3810 & $0.465$ & ACIS-I &  $29.9$ & $16\ 21\ 24.69$ & $+38\ 10\ 08.6$ & $11048$ & $794$  & $22$ & $471$ & $6$ \\
     CL0522-3624 & $0.472$ & ACIS-I &  $26.4$ & $05\ 22\ 15.29$ & $-36\ 25\ 02.7$ & $6871$  & $587$  & $16$  & $440$ & $3$ \\ 
  MACSJ1311.0-0310 & $0.494$ & ACIS-I &  $44.5$ & $13\ 11\ 01.87$ & $-03\ 10\ 39.8$ & $11297$ & $634$  & $25$ & $381$ & $6$  \\
  MACSJ2214.9-1400 & $0.503$ & ACIS-I &  $15.4$ & $22\ 14\ 57.48$ & $-14\ 00\ 09.6$ & $7837$  &  $1318$ & $19$ & $872$ & $5$ \\
  MACSJ0911.2+1746 & $0.505$ & ACIS-I &  $23.0$ & $09\ 11\ 10.61$ & $+17\ 46\ 30.9$ & $4220$  & $1283$ & $16$ & $904$ & $8$ \\
  MACSJ0257.1-2326 & $0.505$ & ACIS-I &  $17.0$ & $02\ 57\ 09.13$ & $-23\ 26\ 04.3$ & $3832$  & $1389$ & $17$ & $478$ & $8$ \\
      V1525+0958 & $0.516$ & ACIS-I &  $28.2$ & $15\ 24\ 40.04$ & $+09\ 57\ 48.9$ & $3613$  &  $575$  & $8$  & $435$ & $4$ \\
   MS0015.9+1609 & $0.541$ & ACIS-I &  $31.0$ & $00\ 18\ 33.36$ & $+16\ 26\ 12.6$ & $9652$  &  $1375$ & $41$ & $913$ & $9$ \\
  CL0848.6+4453 & $0.543$ & ACIS-I & $125.2$ & $08\ 48\ 47.73$ & $+44\ 56\ 13.9$ & $13613$ & $300$ & $5$ & $282$ & $3$ \\ 
  MACSJ1423.8+2404 & $0.543$ & ACIS-S & $105.4$ & $14\ 23\ 47.90$ & $+24\ 04\ 42.2$ & $35182$ & $899$  & $33$ & $603$ & $10$ \\
  MACSJ1149.5+2223 & $0.544$ & ACIS-I &  $51.4$ & $11\ 49\ 35.52$ & $+22\ 23\ 52.7$ & $23253$ & $1470$ & $26$ & $875$ & $8$ \\ 
  MACSJ0717.5+3745 & $0.546$ & ACIS-I &  $74.6$ & $07\ 17\ 31.22$ & $+37\ 45\ 22.6$ & $34326$ & $1389$ & $62$ & $1090$ & $21$ \\
     CL1117+1744 & $0.548$ & ACIS-I &  $37.5$ & $11\ 17\ 29.89$ & $+17\ 44\ 52.1$ & $7098$  & $520$  & $8$  & $500$ & $3$ \\
   MS0451.6-0305 & $0.550$ & ACIS-S &  $37.0$ & $04\ 54\ 11.04$ & $-03\ 00\ 57.8$ & $18100$ & $955$  & $33$ & $486$ & $6$ \\
   MS2053.7-0449 & $0.583$ & ACIS-I &  $35.0$ & $20\ 56\ 21.12$ & $-04\ 37\ 48.4$ & $5428$  & $463$  & $11$  & $293$ & $3$ \\
  MACSJ2129.4-0741 & $0.589$ & ACIS-I &  $18.0$ & $21\ 29\ 25.64$ & $-07\ 41\ 32.0$ & $6226$  & $1055$ & $13$ & $611$ & $5$ \\
  MACSJ0647.7+7014 & $0.591$ & ACIS-I &  $17.9$ & $06\ 47\ 49.95$ & $+70\ 14\ 56.2$ & $5362$  & $1028$ & $20$ & $274$ & $4$ \\
     CL1120+4318 & $0.600$ & ACIS-I &  $18.6$ & $11\ 20\ 07.23$ & $+43\ 18\ 03.6$ & $3452$  & $722$  & $13$  & $599$ & $4$  \\
CLJ0542.8-4100	&	$0.640$	&	ACIS-I	&	$49.9$	&	$05\ 42\ 49.63$ & $-40\ 59\ 56.3$	&	$5026$	&	$744$	&	$12$	&	$771$	&	$4$	\\
LCDCS954	&	$0.670$	&	ACIS-S	&	$26.9$	&	$14\ 20\ 29.25$ & $-11\ 34\ 19.4$	&	$1005$	&	$586$	&	$8$	&	$384$	&	$3$	\\
MACSJ0744.9+3927	&	$0.698$	&	ACIS-I	&	$48.7$	&	$07\ 44\ 52.82$ & $+39\ 27\ 26.1$	&	$9257$		&	$1106$	&	$23$	&	$508$	&	$5$	\\
V1221+4918	&	$0.700$	&	ACIS-I	&	$74.3$	&	$12\ 21\ 25.71$ & $+49\ 18\ 30.4$	&	$2411$	&	$592$	&	$14$	&	$595$	&	$5$	\\
SPT-CL0001-5748	&	$0.700$	&	ACIS-I	&	$29.4$	&	$00\ 00\ 59.91$ & $-57\ 48\ 34.7$	&	$7544$	&	$525$	&	$14$	&	$244$	&	$3$	\\
RCS2327.4-0204	&	$0.704$	&	ACIS-I	&	$73.4$	&	$23\ 27\ 27.68$ & $-02\ 04\ 38.5$	&	$13778$	&	$944$	&	$28$	&	$705$	&	$8$	\\
SPT-CLJ2043-5035 &	$0.720$	&	ACIS-I	&	$76.6$	&	$20\ 43\ 17.48$ & $-50\ 35\ 32.0$	&	$5006$	&	$594$	&	$11$	&	$380$	&	$3$	\\
ClJ1113.1-2615	&	$0.730$	&	ACIS-I	&	$92.5$	&	$11\ 13\ 05.42$ & $-26\ 15\ 39.2$	&	$660$	&	$330$	&	$10$	&	$288$	&	$3$	\\
CLJ2302.8+0844	&	$0.734$	&	ACIS-I	&	$100.6$	&	$23\ 02\ 48.05$ & $+08\ 43\ 49.3$	&	$3649$	&	$627$	&	$10$	&	$350$	&	$3$	\\
SPT-CL2337-5942	&	$0.775$	&	ACIS-I	&	$19.7$	&	$23\ 37\ 24.65$ & $-59\ 42\ 22.7$	&	$2013$	&	$557$	&	$10$	&	$254$	&	$3$	\\
RCS2318+0034	&	$0.780$	&	ACIS-I	&	$112.5$	&	$23\ 18\ 30.88$ & $+00\ 34\ 01.6$	&	$22445$	&	$446$	&	$13$	&	$380$	&	$4$	\\
MS1137.5+6625	&	$0.782$	&	ACIS-I	&	$101.3$	&	$11\ 40\ 22.53$ & $+66\ 08\ 14.3$	&	$3454$	&	$440$	&	$14$	&	$402$	&	$7$	\\
RXJ1350.0+6007	&	$0.810$	&	ACIS-I	&	$55.2$	&	$13\ 50\ 48.18$ & $+60\ 07\ 13.4$	&	$4564$	&	$698$	&	$8$	&	$450$	&	$3$	\\
RXJ1716.9+6708	&	$0.813$	&	ACIS-I	&	$50.7$	&	$17\ 16\ 48.94$ & $+67\ 08\ 25.2$	&	$1180$	&	$418$	&	$9$	&	$481$	&	$3$	\\
EMSS1054.5-0321	&	$0.831$	&	ACIS-S	&	$63.5$	&	$10\ 57\ 00.07$ & $-03\ 37\ 33.1$	&	$3872$	&	$566$	&	$11$	&	$574$	&	$5$	\\
CLJ1226.9+3332	&	$0.888$	&	ACIS-I	&	$29.9$	&	$12\ 26\ 58.07$ & $+33\ 32\ 46.0$	&	$3450$	&	$779$	&	$15$	&	$277$	&	$4$	\\
XMMUJ1230+1339	&	$0.975$	&	ACIS-S	&	$38.4$	&	$12\ 30\ 17.06$ & $+13\ 39\ 08.5$	&	$6538$	&	$344$	&	$9$	&	$287$	&	$4$	\\
J1415.1+3612	&	$1.030$	&	ACIS-S	&	$97.5$	&	$14\ 15\ 11.01$ & $+36\ 12\ 04.1$	&	$8727$	&	$419$	&	$20$	&	$260$	&	$4$	\\
SPT-CL0547-5345	&	$1.067$	&	ACIS-I	&	$28.0$	&	$05\ 46\ 37.25$ & $-53\ 45\ 30.6$	&	$3492$	&	$657$	&	$8$	&	$597$	&	$3$	\\
SPT-CLJ2106-5844	&	$1.132$	&	ACIS-I	&	$47.1$	&	$21\ 06\ 03.38$ & $-58\ 44\ 29.6$	&	$7552$	&	$680$	&	$11$	&	$432$	&	$3$	\\
RDCS1252-2927	&	$1.235$	&	ACIS-I	&	$148.7$	&	$12\ 52\ 54.58$ & $-29\ 27\ 16.9$	&	$13103$	&	$378$	&	$7$	&	$286$	&	$3$	\\
\hline
\end{tabular}

\end{center}
\tablefoot{Columns from left to right list the target name, the adopted redshift, the 
detector used in the observation, the net exposure time (in kilo-seconds) 
after all cleaning processes, the position of the adopted X-ray center in equatorial J2000 coordinates, and
the number of counts measured for each target in the [0.7 - 2] keV band, up to the radial limit $R_\text{out}^\text{spat}$. 
The last four columns list the the upper limit of the radial range investigated in the spatial analysis 
($R_\text{out}^\text{spat}$) and in the spectral analysis ($R_\text{out}^\text{spec}$),
with the number of bins with which we can sample the surface
brightness and the temperature profiles (the temperature bins are obtained by integrating
the spectra between 0.6 and 7 keV).}
\end{table*}

The acquired data are reduced using the CIAO 4.7 software 
\citep[Chandra Interactive Analysis of Observations,][]{fruscione+06} and the calibration database CALDB 4.6.5 (December 2014 release \footnote{\url{http://cxc.harvard.edu/caldb/}}).
This procedure includes a filter for the good time intervals associated with each observation and a correction for the 
charge transfer inefficiency. It removes photons detected in bad CCD columns and pixels, 
it computes calibrated photon energies by applying ACIS gain maps and it corrects for their time dependence. Moreover, it examines the 
background light curves during each observation to detect and remove flaring episodes. We identify bright point sources using the 
\texttt{wavdetect} alogorithm by \citet{vikhlinin+98}, check the results by visual inspection, mask all the detected point sources and 
exclude them from the following analysis.

\section{Spatial and spectral analysis}
\label{sec:analysis}

Obtaining good brightness and temperature profiles is crucial for the quality of the mass estimates.
This strongly depends on the quantity and quality of data obtained for each observation, namely the number of counts measured for the observed 
target and the fraction of counts on the background. 

We extract surface brightness radial profiles from the images in the $[0.7-2]$ keV band, 
by constructing a set of circular annuli around the X-ray emission peak, each one containing at least 100 net source counts.
The background counts are estimated from local regions of the same exposure 
which are free from source emissions (on the same chip as the source region or 
on another chip of the same type used in the observation). 
Following this criterion, we manually select from two to four background regions for each cluster. 
The surface brightness profile is then extracted over an area where the signal-to-noise ratio is always larger than 2, up to the radius $R_\text{out}^\text{spat}$.
In Table~\ref{tab:sample}, we quote the number of counts measured for each target in the $[0.7-2]$ keV band, 
the number of radial bins obtained to sample the surface brightness profile, and $R_\text{out}^\text{spat}$.

For the spectral analysis, we use the CIAO \texttt{specextract} tool to extract the source and the background spectra and to
construct the redistribution matrix files (RMF) and the ancillary response files (ARF) for each annulus. 
The RMF associates to each instrument channel the appropriate photon energy, while the ARF includes information on the effective area, the efficiency 
of the instrument in revealing photons and any additional energy-dependent efficiencies.
The background spectra are extracted from the same background regions used for the spatial analysis. 
The source spectra are extracted from at least three concentric annuli centered on the X-ray surface brightness centroid up to the radius $R_\text{out}^\text{spec}$
where the signal-to-noise is larger than 0.3 in the $[0.6-7]$ keV band. Each spectrum contains at least 500 net source counts in the $[0.6-7]$ keV band. 
For 5 objects (CL0848.6+4453, LCDCS954, CLJ1113.1-2615, CLJ2302.8+0844 and RDCS1252-2927), we consider a minimum of 200 net counts to resolve 
the temperature profile in 3 independent radial bins.
In Table~\ref{tab:sample}, we also report the radial limit probed in the spectral analysis ($R_\text{out}^\text{spec}$) 
and the number of bins with which we can sample the temperature profiles by integrating the spectra between 0.6 and 7 keV.

For each annulus, the spectrum is analysed with the X-ray spectral fitting software XSPEC \citep{arnaud96}.  
We adopt a collisionally-ionized diffuse gas emission model (\texttt{apec}), multiplied by an absorption component (\texttt{tbabs}). 
In this model, we fix the redshift to the value obtained from the optical spectroscopy and the absorbing equivalent hydrogen column 
density $N_H$ to the value of the Galactic absorption inferred from radio HI maps in \citet{dl90}.
Then, the free parameters in the spectral fitting model are the emission-weighted temperature, the metallicity and the normalisation of the thermal spectrum. 
The fit is performed in the energy range $[0.6-7]$ keV applying the Cash statistics \citep{cash79} as implemented in XSPEC. 
The Cash statistics is a maximum-likelihood estimator based on the Poisson distribution of the detected source plus background counts
and is preferable for low signal-to-noise spectra (e.g. Nousek \& Shue 1989).

The gas density profile is then obtained through the geometrical deprojection (e.g. Fabian et al. 1981, Ettori et al. 2002) 
of both the surface brightness profile $S_\text{b}$ and the normalisation $K$ of the thermal model fitted in the spectral analysis.

\section{The hydrostatic mass profile}
\label{sec:methods}
The total mass of X-ray luminous galaxy clusters can be estimated from the observed gas density $n_\text{gas}$ and temperature $T_\text{gas}$ profiles. 
The Euler's equation for a spherically-symmetric distribution of gas with pressure $P_\text{gas}$ and density $\rho_\text{gas}$, 
in hydrostatic equilibrium with the underlying gravitational potential $\phi$, requires \citep{bt87}:

\begin{equation}
\label{eq:hee0}
\frac{1}{\rho_\text{gas}} \frac{dP_\text{gas}}{dr} = - \frac{d\phi}{dr} = -\frac{GM_\text{tot}(<r)}{r^2} \,,
\end{equation}
which is better known as the hydrostatic equilibrium equation (HEE).
Solving equation~(\ref{eq:hee0}) for the total mass $M_\text{tot}$ and considering the perfect gas law, 
$P_\text{gas} = \rho_\text{gas}\,kT_\text{gas}/(\mu m_\text{p}) = n_\text{gas}\,kT_\text{gas}$, 
we can obtain the total mass of the clusters as a function of our observables, the gas density and temperature profiles \citep[see e.g.][for a recent review]{ettori+13}:

\begin{equation}
\label{eq:hee}
M_\text{tot}(<r)=-\frac{kT_\text{gas}(r)r}{\mu m_\text{p} G} \left( \frac{d\ln n_\text{gas}}{d\ln r}+\frac{d\ln T_\text{gas}}{d\ln r} \right) \,.
\end{equation}
Here, $G$ is the gravitational constant, $k$ is the Boltzmann's constant, $m_\text{p}$ is the proton mass, $\mu=0.6$ is the mean molecular weight 
of the gas and $n_\text{gas}=\rho_\text{gas}/\mu m_\text{p}$ is the sum of the electron and the ion densities.

We consider a galaxy cluster as a spherical region with a radius $R_\Delta$, where $\Delta$ is the mean overdensity 
with respect to the critical density of the Universe at the cluster's redshift and we will define all the quantities describing the cluster's 
mass profile in relation to the overdensity $\Delta = 200$. 
We will define the masses with respect to the critical density of the Universe. \citet{dk15} have pointed out that the 
time evolution of the concentration with the peak height $\nu$ exhibits the smallest deviations from universality if this definition 
is adopted.

\begin{figure}[h!]
\vbox{
 \includegraphics[width=0.5\textwidth]{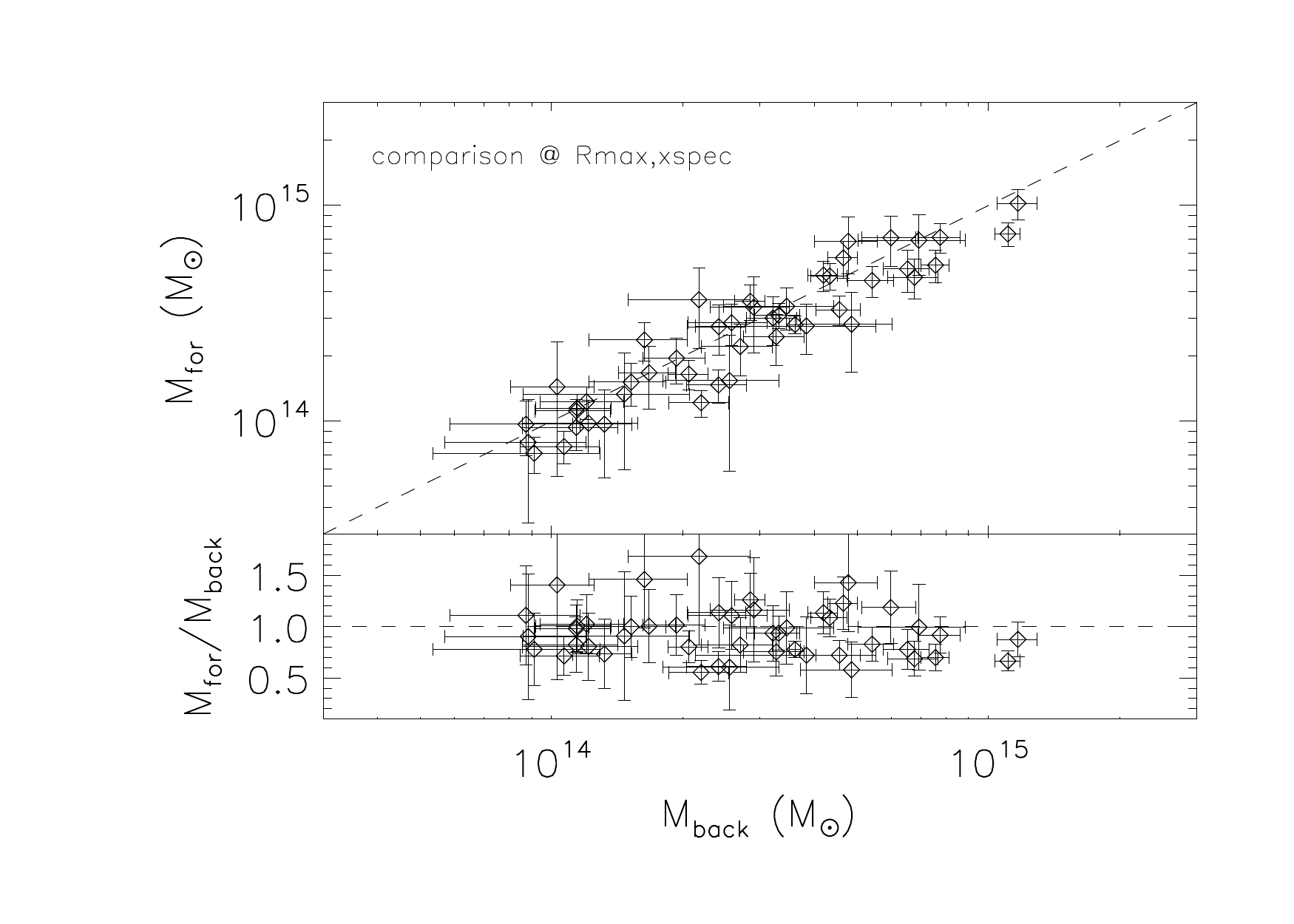}
 \includegraphics[width=0.5\textwidth]{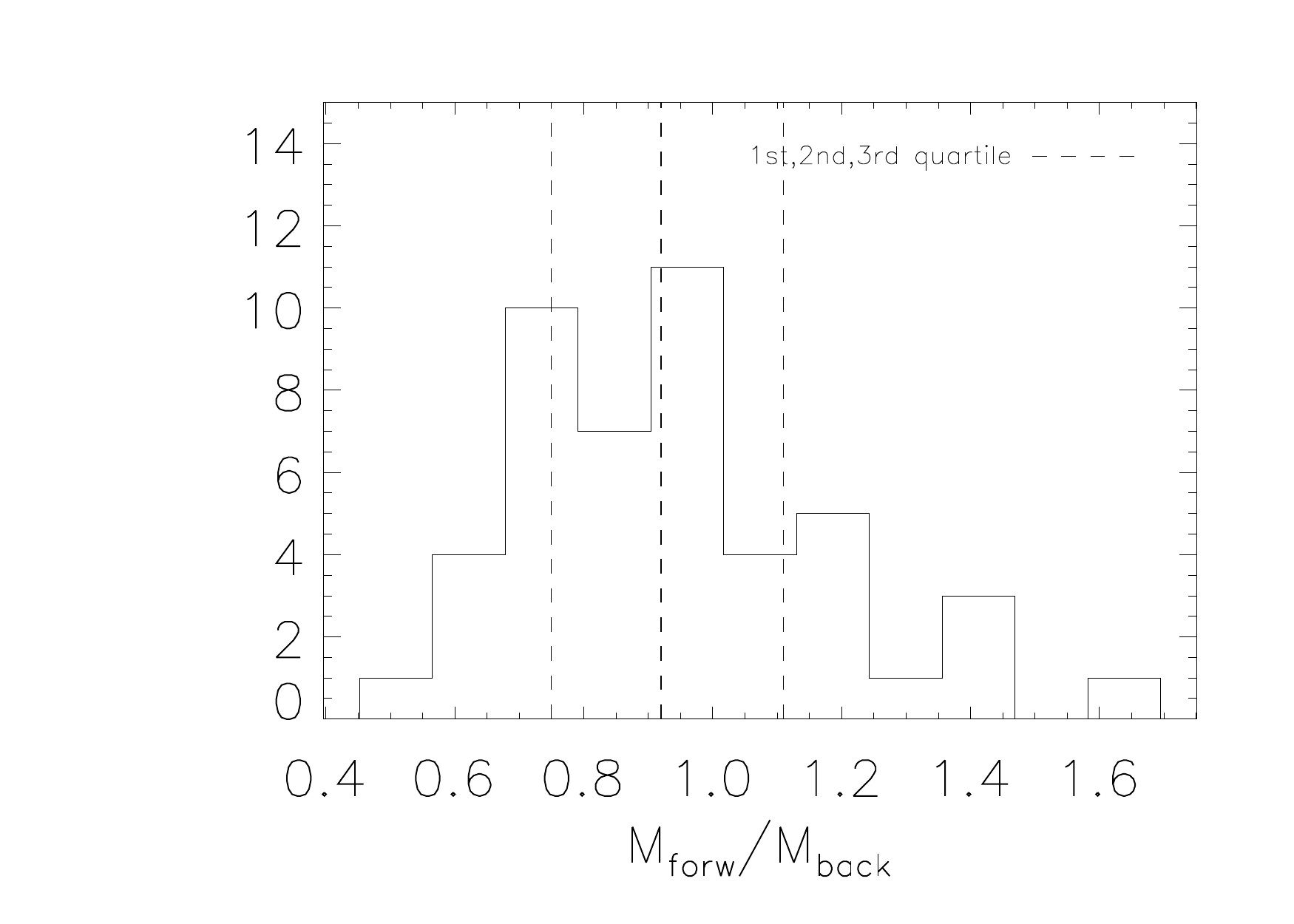}
 \includegraphics[width=0.5\textwidth]{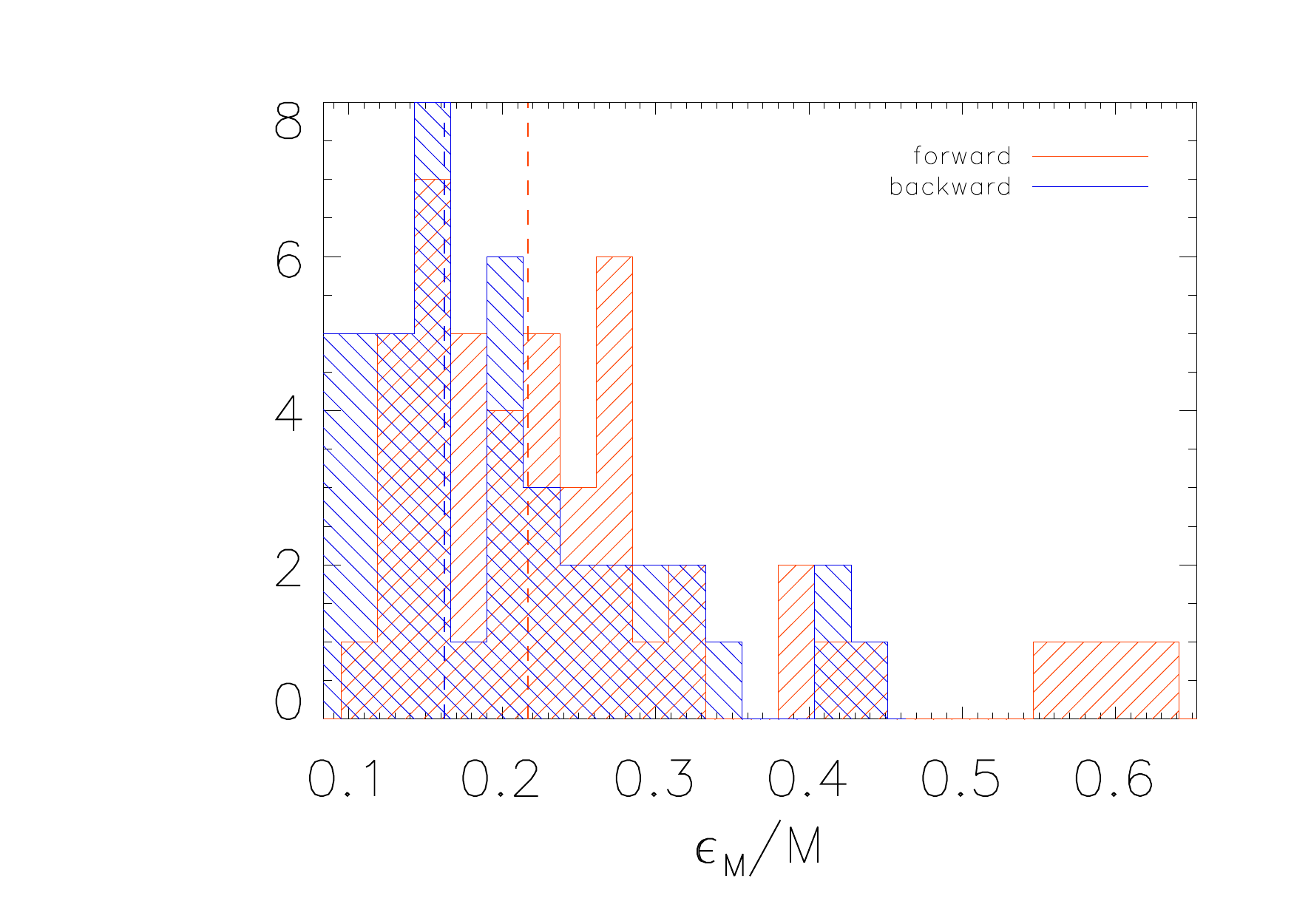}
 }
\caption{
{\it Top:} comparison between mass estimates obtained following the \for\ ($M_\text{for}$) and \back\ ($M_\text{back}$) for the 47 clusters 
of our sample. The lower panel shows the $M_\text{for}/M_\text{back}$ ratio of individual clusters against $M_\text{back}$. 
The dashed line shows the one-to-one relation.
The comparison is made at the outermost radius measured in the spectral analysis for each cluster.
{\it Middle:} distribution of the mass ratios. 
{\it Bottom:} distribution of the relative errors.
}
\label{fig:mfmb}
\end{figure}

As described in \citet{ettori+13}, equation~(\ref{eq:hee}) can be solved at least with two different approaches, adopting either a \back\ or a \for. 

The \back\ follows the approach described in \citet{ettori+10}. Briefly, it consists in adopting a functional form to describe the 
total mass profile, while there is no parametrisation of the gas temperature and density profiles. 
In this work, we adopt the NFW profile, so that:
\begin{eqnarray}
\lefteqn{ M_\text{tot}(<r)=4\pi r_\text{s}^3 \rho_\text{s} f(x), } \nonumber \\
\lefteqn{ \rho_\text{s}=\rho_\text{c,z} \frac{200}{3} \frac{c^3}{\ln(1+c)-c/(1+c)} } , \nonumber \\
\lefteqn{ f(x)=\ln(1+x)-\frac{x}{1+x} } ,
\end{eqnarray}
where $x=r/r_\text{s}$. 
This model is a function of two parameters: the scale radius $r_\text{s}$ and the concentration $c$, which are related 
by the relation $R_{200}=c_{200}\times r_\text{s}$. The best-fit parameters are searched over a grid of values in the $(r_\text{s},c)$ 
plane and they are constrained by minimising the following $\chi^2$ statistics:
\begin{equation}
\chi^2_T=\sum_i \frac{(T_\text{data,i}-T_\text{model,i})^2}{\epsilon^2_\text{T,i}}\,,
\end{equation}
where the sum is done over the annuli of the spectral analysis; $T_\text{data}$ are the temperature measurements obtained in the spectral 
analysis; $T_\text{model}$ are the values obtained by projecting the estimates of $T_\text{gas}$ (recovered from the inversion of the HEE eq.
(\ref{eq:hee}) for a given gas density and total mass profiles) over the annuli used in the spectral analysis, according to \citet{mazzotta+04}; 
$\epsilon_\text{T}$ is the error on the spectral measurements.
The search for the minimum in the $\chi^2_T$ distribution proceeds, first, in identifying a minimum over a grid of $50\times50$ points in which
the range of the two free parameters ($50$ kpc $< r_\text{s} < \max(R_\text{out}^\text{spat},\,R_\text{out}^\text{spec})$; $0.2<c<20$)
is divided regularly.
Then, we obtain the refined best-fit values for the $(r_\text{s},c)$ parameters, looking for a minimum over a $100\times100$ grid in a $5\sigma$ range around the first identified minimum.
Considering the strong correlation present between the free parameters, and to fully represent their probability distribution, we estimate, and quote in Table~\ref{tab:cm},
the probability weighted means of the concentration $c_{200}$ and of the mass $M_{200}$.
The mass is obtained as $200 \, \rho_\text{c,z} \, 4/3 \pi R_{200}^3$, where $R_{200} = r_\text{s} \times c_{200}$ and propagates the joint probability distribution evaluated
for the grid of values of the $(r_\text{s},c)$ parameters.
In Table~\ref{tab:cm}, we quote the best-fit results for $c_{200}$ and $M_{200}$ derived from the \back. 

In the \for\, some parametric functions are used to model the three-dimensional gas density and temperature radial profiles. 
This is similar to what is described in e.g. \citet{vikhlinin+06}, where the adopted functional forms are projected along the line of sight to fit the observed projected quantities.
In the present analysis, we model directly the deprojected three-dimensional profiles.
The gas density distribution is parametrised by a double $\beta$-model: 
\begin{equation}
n_\text{gas}(r)=\frac{n_0}{[1+(r/r_0)^2]^{1.5\alpha}} + \frac{n_1}{[1+(r/r_1)^2]^{1.5\beta}}
\end{equation}
where $n_0, n_1, r_0, r_1, \alpha, \beta$ are the free parameters of the model. 
The three-dimensional temperature profile is modelled as
\begin{equation}
T(r)=T_0 \frac{a+(r/r_\text{in})^b}{[1+(r/r_\text{in})^b][1+(r/r_\text{out})^2]^d} \,,
\end{equation}
where $T_0, r_\text{in}, r_\text{out}, a, b, d$ are the free parameters of the model. 
These profiles, with their best-fit values and intervals, are then used to recover the mass profile through equation~(\ref{eq:hee}). 

The two methods show a good agreement between the two estimates of the mass contained within the outermost radius measured in the 
spectral analysis, as shown in Fig. \ref{fig:mfmb}.
In fact, the ratio between the two mass estimates has a median (1st, 3rd quartile) value of $0.92$ ($0.75$, $1.11$).
The distributions of the relative errors are also quite similar, with a median value of $22\%$ for the \for\ and $16\%$ for the \back\ .
For the following analysis, we have chosen to follow the \back\ since it requires only two parameters 
and provides more reliable estimates of the uncertainties \citep[see e.g.][]{ma11}.

Eleven clusters of our sample are among the targets of the CLASH (Cluster Lensing and Supernova survey with Hubble) program \citep{pos+al12}.
CLASH was a Hubble Multi-Cycle Treasury program with the main science goal to obtain well-constrained gravitational-lensing mass profiles for a sample of 25 massive galaxy clusters in the redshift range $0.2-0.9$. Twenty of these clusters were selected to have relatively round X-ray isophotes centered on a prominent Brightest Central Galaxy. The remaining five were chosen for their capability to provide extraordinary signal for gravitational lensing.
\citet{donahue+14} derive the mass profiles of the CLASH clusters from X-ray observations (either \chandra\ or \xmm) in order to compare them
with lensing results. We compare the masses at the radius $R_{500}$ listed in their Table 4 for the \chandra\ data with the masses derived 
from our {\it backward} analysis, calculated at the same physical radius. \citet{donahue+14} invoke the HEE as we do, but they reconstruct
the mass profiles in a different way. They use the JACO (Joint Analysis of Clusters Observations) fitting tool \citep{mahdavi+07} which employs
parametric models for both the dark matter and the gas density profiles (a NFW model and a combination of $\beta-$models, respectively, 
in this case), under the assumption of a spherically-symmetric ICM in hydrostatic equilibrium with the DM potential, to reconstruct the 
projected spectra in each annular bin that are then jointly fitted to the observed events to constrain the model parameters. 
We find an encouraging agreement between the two outcomes. The median (1st, 3rd quartile)
of the $M_\text{back}/M_\text{CLASH}$ distribution for the 11 shared clusters is $1.09$ ($0.86$, $1.44$). The distributions of the relative 
errors provided by the two analyses are also comparable, with a median value of $21\%$ for our \back\ and $26\%$ for the method employed by 
\citet{donahue+14}.

\begin{table*}
\caption{Results on the mass reconstruction.} 
\label{tab:cm}
\vspace*{-0.2cm}
\begin{center}
\begin{tabular}{c c c c c c c }
\hline\hline
name & z & kT [keV] & $M_{gas,500} [10^{14} M_\odot]$ &  $c_{200}$ & $M_{200} [10^{14} M_\odot]$ & $\chi^2/dof\,(P)$ \\ 
\hline
  MACSJ0159.8-0849 & $0.405$ & $9.2 \pm 0.6$ & $1.29 \pm 0.07$ &  $4.3 \pm 0.8$ & $17.8 \pm  5.4$ & $1.42\,(0.84)$ \\
  MACSJ2228.5+2037  & $0.412$ & $9.4 \pm 0.7$ & $1.57 \pm 0.12$ &  $2.7 \pm 1.1$ & $15.6 \pm  5.1$ & $0.09\,(0.04)$ \\
   MS1621.5+2640   & $0.426$ & $6.7 \pm 0.6$ & $0.95 \pm 0.06$ &  $2.4 \pm 0.9$ & $13.0 \pm  4.1$ & $0.82\,(0.52)$ \\
  MACSJ1206.2-0848  & $0.440$ & $12.5 \pm 1.0$ & $2.25 \pm 0.10$ &  $2.5 \pm 0.5$ & $38.1 \pm  10.3$ & $1.62\,(0.82)$ \\
  MACSJ2243.3-0935   & $0.447$ & $8.4 \pm 0.6$ & $1.73 \pm 0.11$ &  $2.7 \pm 1.2$ & $14.5 \pm  4.4$ & $1.29\,(0.73)$ \\
  MACSJ0329.7-0211  & $0.450$ & $7.7 \pm 0.6$ & $1.00 \pm 0.06$ &  $3.5 \pm 0.7$ & $15.9 \pm  5.6$ & $0.62\,(0.29)$ \\
  RXJ1347.5-1145   & $0.451$ & $15.1 \pm 0.8$ & $2.43 \pm 0.11$ &  $4.5 \pm 0.6$ & $40.1 \pm 11.2$ & $0.73\,(0.33)$ \\
      V1701+6414   & $0.453$ & $6.3 \pm 0.7$ & $0.78 \pm 0.07$ & $2.2 \pm 1.0$ & $7.8 \pm 3.4$ & $0.86\,(0.52)$ \\
  MACSJ1621.6+3810 & $0.465$ & $9.1 \pm 1.0$ & $0.81 \pm 0.05$ &  $3.4 \pm 1.0$ & $21.7 \pm 10.9$ & $0.62\,(0.35)$ \\
     CL0522-3624   & $0.472$ & $4.2 \pm 1.2$ & $0.23 \pm 0.03$ &  $6.3 \pm 4.9$ & $ 6.1 \pm  4.6$ & $0.02\,(0.11)$ \\
  MACSJ1311.0-0310 & $0.494$ & $5.7 \pm 0.4$ & $0.77 \pm 0.03$ &  $2.6 \pm 0.8$ & $18.6 \pm  7.8$ & $0.52\,(0.28)$ \\
  MACSJ2214.9-1400   & $0.503$ & $11.9 \pm 1.6$ & $1.41 \pm 0.13$ &  $4.4 \pm 2.9$ & $17.9 \pm  9.0$ & $0.59\,(0.38)$ \\
  MACSJ0911.2+1746 & $0.505$ & $7.9 \pm 1.0$ & $1.12 \pm 0.74$ &  $2.5 \pm 1.0$ & $15.5 \pm  5.1$ & $1.16\,(0.68)$ \\
  MACSJ0257.1-2326  & $0.505$ & $8.6 \pm 0.9$ & $1.31 \pm 0.13$ &  $3.9 \pm 2.3$ & $17.3 \pm  8.7$ & $0.63\,(0.29)$ \\
      V1525+0958   & $0.516$ & $4.7 \pm 0.7$ & $0.52 \pm 0.03$ &  $2.5 \pm 1.3$ & $11.1 \pm  5.5$ & $1.05\,(0.65)$ \\
   MS0015.9+1609   & $0.541$ & $9.9 \pm 0.8$ & $1.78 \pm 0.12$ &  $2.3 \pm 0.6$ & $19.9 \pm  5.1$ & $0.92\,(0.51)$ \\
   CL0848.6+4453   & $0.543$ & $4.9 \pm 0.8$ & $0.16 \pm 0.01$ &  $5.2 \pm 4.3$ & $9.4 \pm 8.6$ & $0.11\,(0.26)$ \\
  MACSJ1423.8+2404 & $0.543$ & $7.5 \pm 0.3$ & $0.70 \pm 0.03$ & $6.2 \pm 0.4$ & $ 7.8 \pm  0.8$ & $1.44\,(0.83)$ \\ 
  MACSJ1149.5+2223  & $0.544$ & $10.8 \pm 0.7$ & $1.73 \pm 0.10$ &  $3.3 \pm 2.0$ & $13.3 \pm  4.5$ & $0.85\,(0.47)$ \\
  MACSJ0717.5+3745  & $0.546$ & $7.9 \pm 0.5$ & $2.52 \pm 0.12$ &  $3.6 \pm 0.9$ & $21.7 \pm  4.0$ & $1.52\,(0.93)$ \\
     CL1117+1744   & $0.548$ & $2.5 \pm 1.2$ & $0.19 \pm 0.02$ &  $4.8 \pm 4.5$ & $ 2.2 \pm  1.6$ & $0.45\,(0.50)$ \\
   MS0451.6-0305   & $0.550$ & $11.2 \pm 0.7$ & $1.78 \pm 0.06$ &  $3.2 \pm 1.4$ & $28.5 \pm 11.3$ & $1.30\,(0.73)$ \\ 
   MS2053.7-0449   & $0.583$ & $5.6 \pm 1.6$ & $0.36 \pm 0.03$ &  $4.3 \pm 3.7$ & $ 8.1 \pm  6.0$ & $0.40\,(0.47)$ \\
  MACSJ2129.4-0741   & $0.589$ & $11.6 \pm 2.1$ & $1.23 \pm 0.08$ &  $6.5 \pm 4.4$ & $16.0 \pm  9.6$ & $0.91\,(0.57)$ \\
  MACSJ0647.7+7014   & $0.591$ & $13.2 \pm 2.5$ & $1.74 \pm 0.12$ &  $3.7 \pm 2.4$ & $25.6 \pm 15.2$ & $0.62\,(0.46)$ \\
     CL1120+4318   & $0.600$ & $4.9 \pm 1.4$ & $0.65 \pm 0.09$ &  $4.7 \pm 4.0$ & $ 7.0 \pm  4.2$ & $1.03\,(0.65)$ \\
  CLJ0542.8-4100   & $0.640$ & $6.0 \pm 0.8$ & $0.43 \pm 0.03$ &  $7.0 \pm 5.2$ & $ 6.5 \pm  3.6$ & $1.13\,(0.68)$ \\
        LCDCS954   & $0.670$ & $3.9 \pm 0.8$ & $0.17 \pm 0.02$ &  $4.8 \pm 4.5$ & $ 2.2 \pm  1.7$ & $1.75\,(0.81)$ \\
  MACSJ0744.9+3927 & $0.698$ & $9.0 \pm 0.7$ & $1.05 \pm 0.07$ &  $6.2 \pm 2.8$ & $ 9.7 \pm  4.9$ & $1.08\,(0.65)$ \\
      V1221+4918   & $0.700$ & $6.3 \pm 0.8$ & $0.40 \pm 0.03$ &  $6.1 \pm 4.8$ & $ 6.6 \pm  4.3$ & $1.46\,(0.78)$ \\
 SPT-CL0001-5748   & $0.700$ & $6.5 \pm 1.0$ & $0.52 \pm 0.03$ &  $5.1 \pm 3.3$ & $13.3 \pm 11.4$ & $0.25\,(0.38)$ \\
  RCS2327.4-0204   & $0.704$ & $9.8 \pm 0.5$ & $1.66 \pm 0.07$ &   $2.2 \pm 0.4$ & $31.3 \pm  7.7$ & $0.72\,(0.37)$ \\
SPT-CLJ2043-5035   & $0.720$ & $6.5 \pm 1.1$ & $0.98 \pm 0.06$ &  $2.6 \pm 1.3$ & $15.1 \pm  8.1$ & $0.18\,(0.33)$ \\
  ClJ1113.1-2615   & $0.730$ & $3.9 \pm 0.7$ & $0.17 \pm 0.02$ &  $6.0 \pm 4.4$ & $ 8.1 \pm  6.8$ & $0.68\,(0.59)$ \\
  CLJ2302.8+0844   & $0.734$ & $11.4 \pm 2.9$ & $0.38 \pm 0.04$ &  $3.2 \pm 2.9$ & $ 7.9 \pm  5.0$ & $3.29\,(0.93)$ \\
 SPT-CL2337-5942   & $0.775$ & $9.3 \pm 1.7$ & $1.14 \pm 0.06$ &  $4.8 \pm 3.8$ & $21.2 \pm 14.1$ & $0.05\,(0.18)$ \\
    RCS2318+0034   & $0.780$ & $10.4 \pm 2.2$ & $0.80 \pm 0.03$ &  $4.8 \pm 3.7$ & $22.9 \pm 17.3$ & $0.27\,(0.24)$ \\
   MS1137.5+6625   & $0.782$ & $5.2 \pm 0.4$ & $0.48 \pm 0.03$ &  $3.6 \pm 1.9$ & $15.2 \pm  8.8$ & $2.07\,(0.93)$ \\
  RXJ1350.0+6007   & $0.810$ & $4.0 \pm 0.6$ & $0.22 \pm 0.03$ &  $5.0 \pm 4.5$ & $ 2.8 \pm  1.5$ & $0.02\,(0.12)$ \\
  RXJ1716.9+6708   & $0.813$ & $4.7 \pm 0.8$ & $0.28 \pm 0.02$ &  $6.6 \pm 5.3$ & $ 6.5 \pm  4.9$ & $1.75\,(0.81)$ \\ 
 EMSS1054.5-0321   & $0.831$ & $11.1 \pm 1.2$ & $1.15 \pm 0.03$ &  $3.8 \pm 3.2$ & $16.3 \pm  8.8$ & $0.71\,(0.45)$ \\
  CLJ1226.9+3332   & $0.888$ & $14.3 \pm 2.4$ & $1.66 \pm 0.10$ &  $4.2 \pm 2.9$ & $33.7 \pm 21.2$ & $0.11\,(0.11)$ \\
  XMMUJ1230+1339   & $0.975$ & $4.3 \pm 1.1$ & $0.37 \pm 0.03$ &  $4.2 \pm 3.7$ & $ 8.7 \pm  7.1$ & $0.35\,(0.30)$ \\
    J1415.1+3612   & $1.030$ & $6.2 \pm 0.7$ & $0.34 \pm 0.02$ &  $3.3 \pm 2.5$ & $10.0 \pm  6.9$ & $0.71\,(0.51)$ \\
 SPT-CL0547-5345   & $1.067$ & $6.9 \pm 1.8$ & $0.58 \pm 0.07$ &  $6.0 \pm 4.7$ & $11.9 \pm  8.8$ & $0.20\,(0.34)$ \\
 SPT-CLJ2106-5844   & $1.132$ & $8.9 \pm 1.2$ & $1.23 \pm 0.06$ &  $4.9 \pm 4.5$ & $ 9.0 \pm  5.4$ & $1.57\,(0.79)$ \\
     RDCS1252-2927   & $1.235$ & $3.7 \pm 1.0$ & $0.22 \pm 0.03$ &  $4.6 \pm 3.9$ & $ 5.6 \pm  4.5$ & $0.25\,(0.38)$ \\
\hline
\end{tabular}
\end{center}
\tablefoot{Columns from left to right list the target name, the adopted redshift, the mean spectral gas temperature, the gas mass within $R_{500}$, 
the probability weighted mean of the mass concentration and of the mass within $\Delta=200$ obtained as described in Sect.~\ref{sec:methods}, 
and the $\chi^2$ divided by the degrees of freedom (i.e. the number of temperature bins listed in the last column of Table \ref{tab:sample} minus two) 
and the corresponding probability that a random variable from a $\chi^2$ distribution with a given degrees of freedom is less or equal to the observed $\chi^2$ value.
}
\end{table*}

\subsection{Comments on the best-fit parameters}

\begin{figure*}
\centering
  \includegraphics[width=0.7\textwidth]{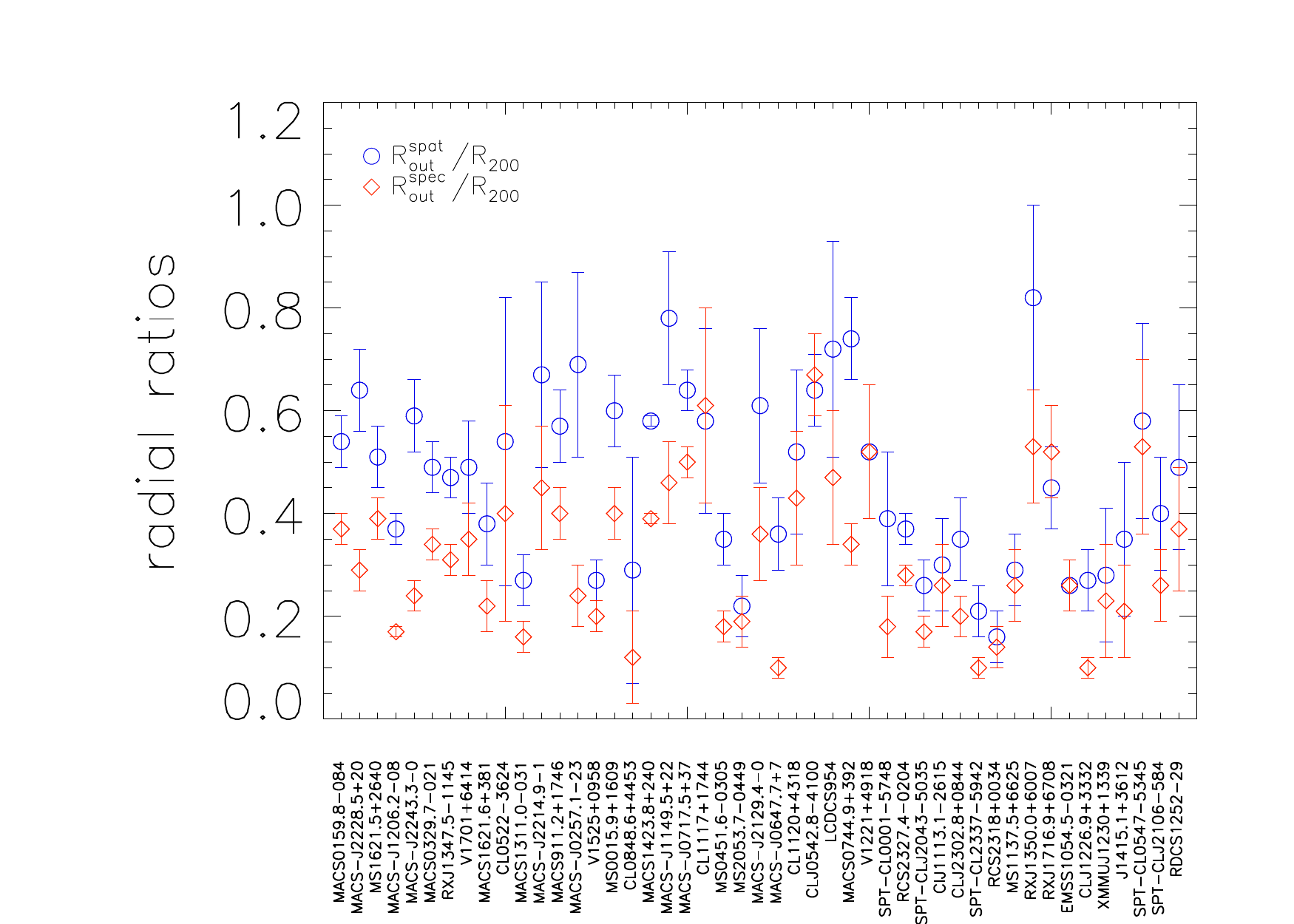}
  \hbox{
  \includegraphics[width=0.4\textwidth]{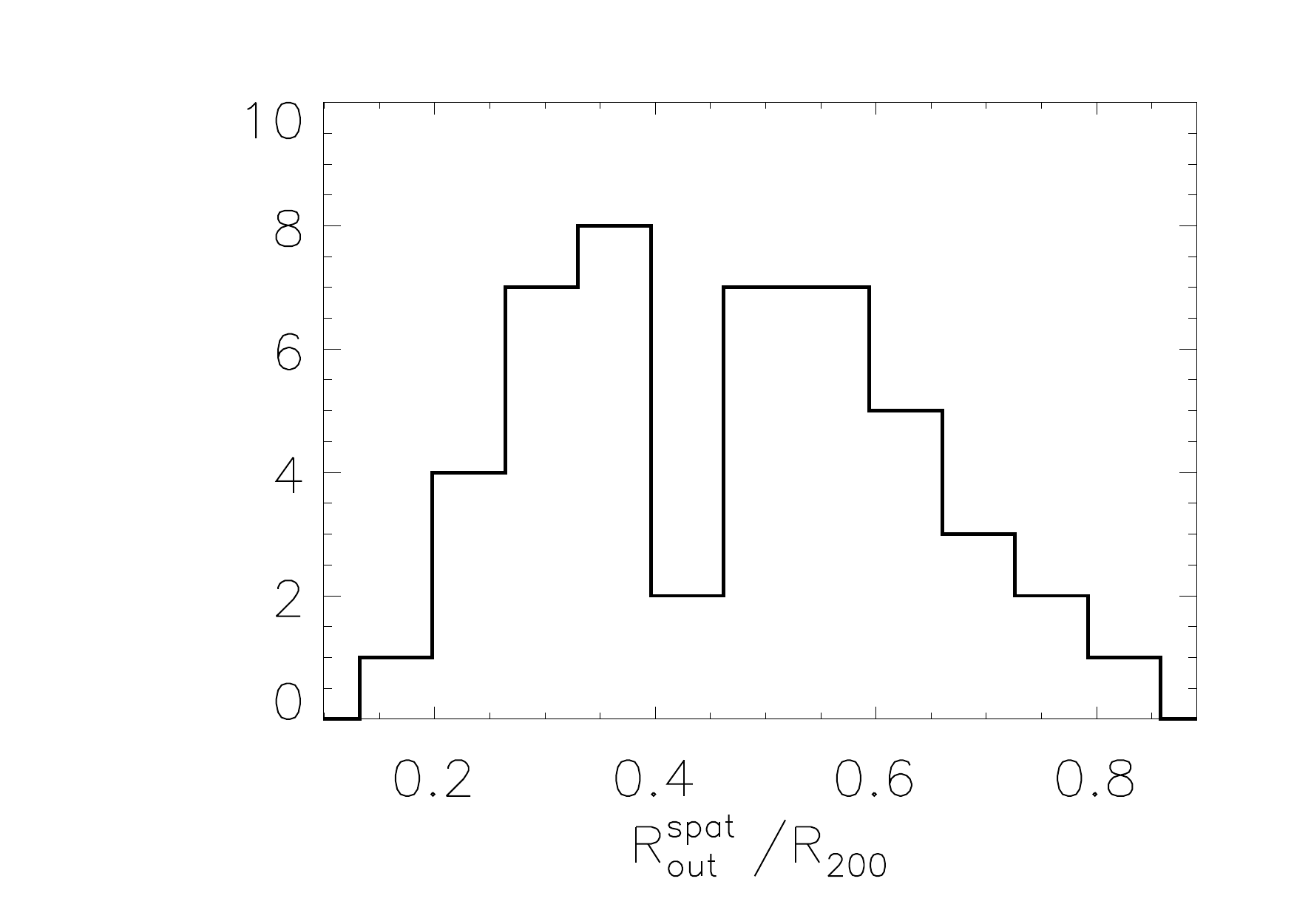}
  \includegraphics[width=0.4\textwidth]{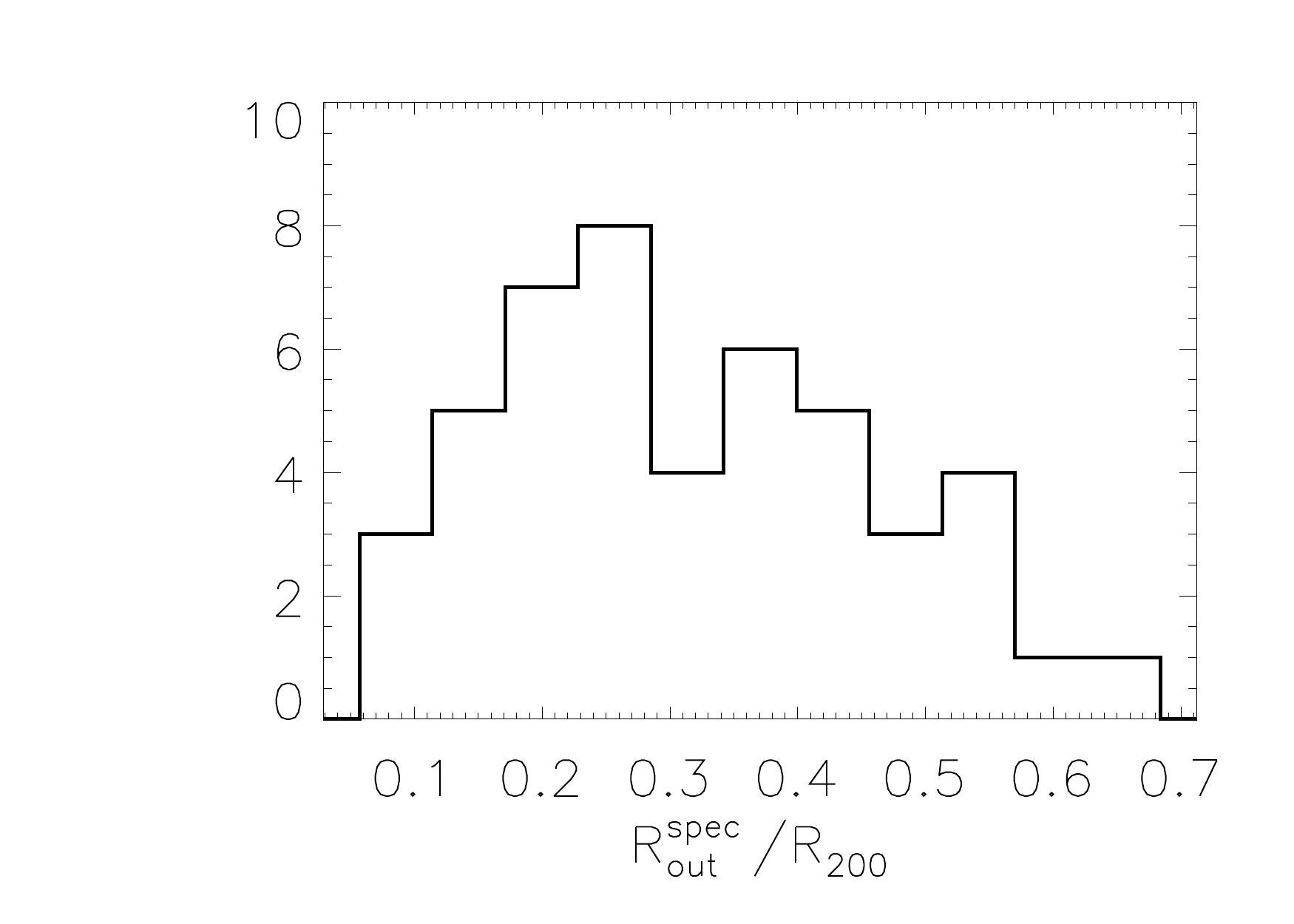}
  }
\caption{{\it Top:} For each cluster in the final sample, we show: the ratio between the upper limit of the radial range investigated in the 
spatial analysis and our estimate of $R_{200}$ (blue circles); the ratio between the maximum radial extension of the spectral analysis and 
$R_{200}$ (red diamonds). 
{\it Bottom:} distributions of the radial ratios.}
\label{fig:radialextent}
\end{figure*}

The radial extension probed with our X-ray measurements span a typical range 35 kpc $\la R_\text{spat} \la$ 700 kpc and 
65 kpc $\la R_\text{spec} \la$ 480 kpc for the spatial and spectral analyses, respectively.
We will use the results on the $c-M$ relation estimated at $R_{200}$ in order to allow a direct comparison with the predictions from simulations. 
In order to check the significance of our estimates, we compare for each cluster
our estimates of $R_{200}$ with the upper limit of the radial range investigated in the spatial and in the spectral analyses. 
The results are shown in Fig. \ref{fig:radialextent}, where we also show the distributions of each of the ratios investigated.

As usual in the X-ray analysis, the estimate of $R_{200}$ exceeds the radial extension of the spatial and the spectral analyses in almost all cases.
For the $R_\text{out}^\text{spat}/R_{200}$ ratio, we measure a median value (1st, 3rd quartiles) of $0.49\,(0.30,\,0.59)$ and a median 
relative dispersion of $21\%$, while we obtain $0.29\,(0.20,\,0.40)$ and a median relative dispersion of $20\%$ for the $R_\text{out}^\text{spec}/R_{200}$ ratio.

This means that we are not able to sample directly our objects up to $R_{200}$ in both the surface brightness and the temperature profiles, 
as expected given that both the observational strategy and the background characterisation were not optimised to this purpose \citep[see e.g.][]{em11}.

However, $R_{200}$ is here treated as a quantity derived from the best-fit parameters of our procedure for the assumed mass model 
($R_{200} = r_\text{s} \times c_{200}$), and does not imply a direct extrapolation of the mass profile to recover it.

More interesting is to consider the goodness of the fitting procedure. 
As we quote in the last column of Table~\ref{tab:cm}, the NFW model provides a reasonable
description of the cluster gravitational potential for all our clusters. The probability that a random variable from a $\chi^2$ distribution 
with a given degree of freedom is less or equal to the observed $\chi^2$ value is 50\% (median of the observed distribution) 
\footnote {We remind that a reduced $\chi^2$ of 1 would have an associated probability of 68.3\% for a degree-of-freedom of 1 and of 51.9\% for d.o.f.=100.}.
We have only one object with a very low probability ($<$5\%; MACSJ2228.5+2037) that suggests an over-estimate of the error bars, and none with a probability larger than 95\%.
Nonetheless, deviations are expected in a sample of about 50 clusters and this object has also been considered in the following analysis.

\subsection{Comparison with lensing estimates}
A useful test for the reliability of our hydrostatic mass estimates is the comparison with results from lensing. The $LC^2$-\texttt{single} 
catalogue is a collection of 506 galaxy clusters from literature with mass measurements based on weak lensing \citep{ser15_comalit_III}. 
Cluster masses in $LC^2$-\texttt{single} are uniformed to our reference cosmology. By cross-matching with the $LC^2$ 
catalogue\footnote{We use the \texttt{LC2-single\_v2.0.dat} version publicly available at \url{http://pico.bo.astro.it/~sereno/CoMaLit/LC2/}.} 
we find that 32 out of 47 clusters of our sample have weak lensing reconstructed mass.

\begin{figure*}
\hbox{
  \includegraphics[width=0.5\textwidth]{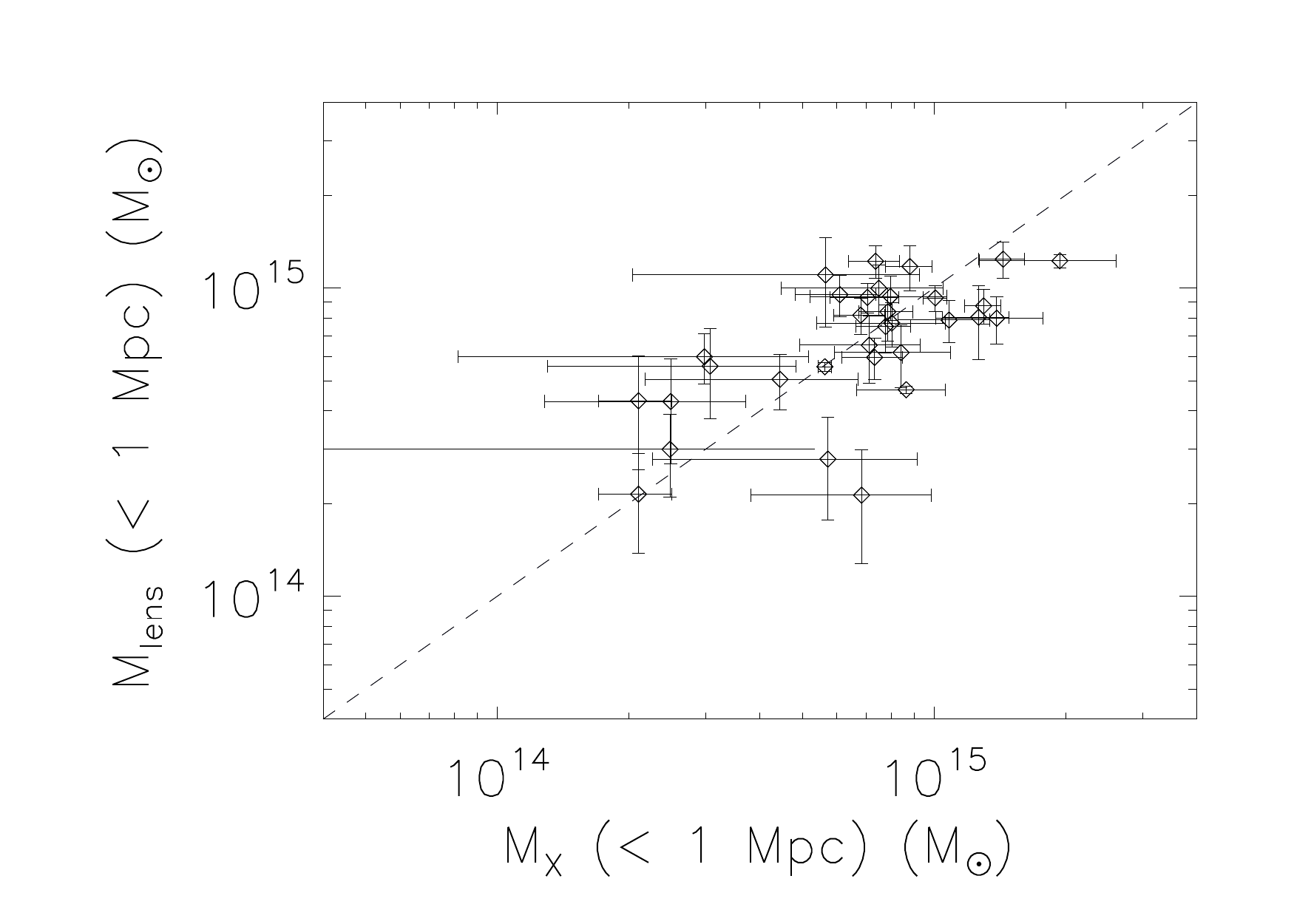}
  \includegraphics[width=0.5\textwidth]{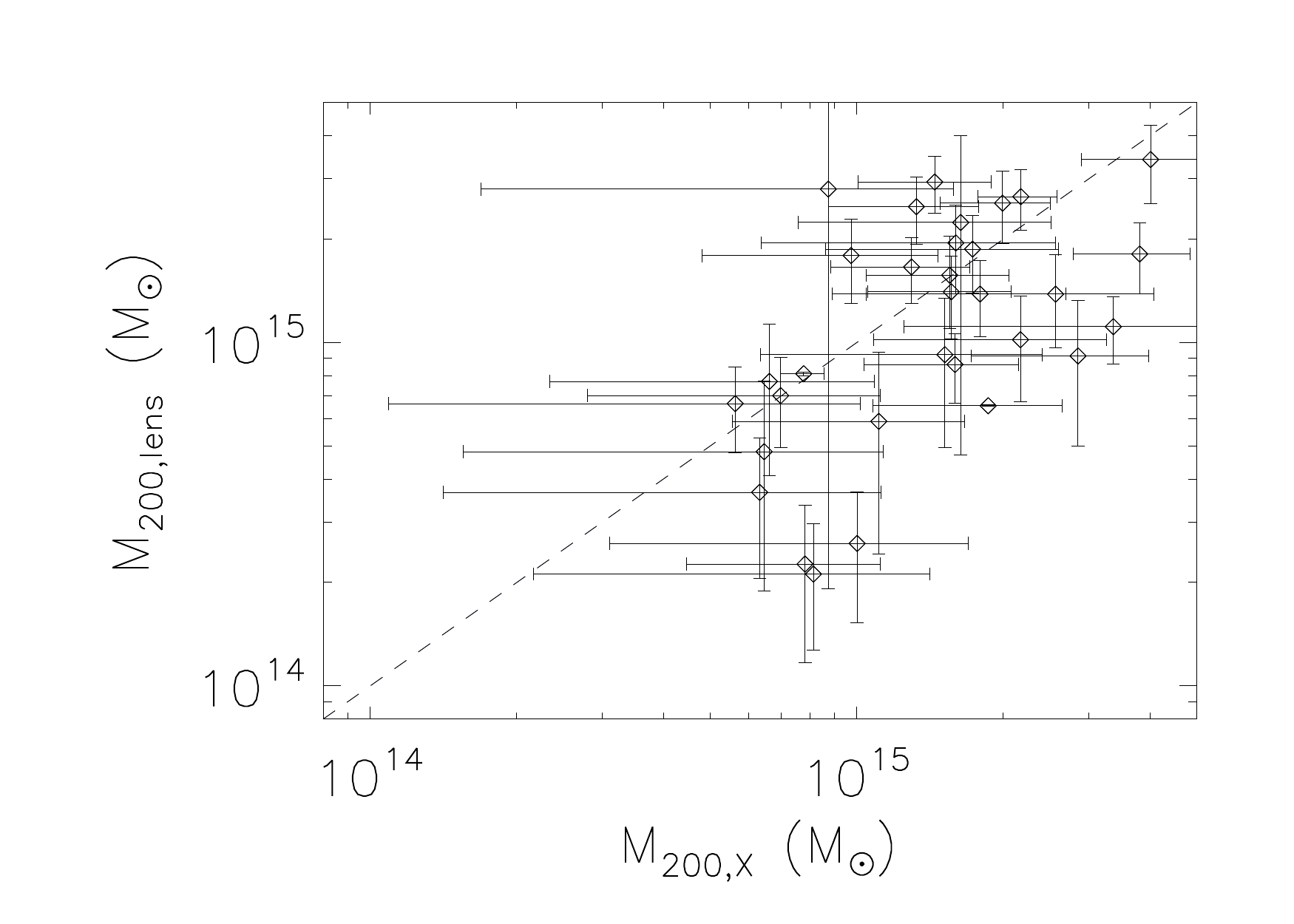}
} 
\caption{Comparison on the mass estimates within $1$ Mpc ({\it left}) and $R_{200}$ ({\it right}) for the objects in common between our sample
of X-ray measurements and the ones available in the lensing $LC^2$-\texttt{single} catalogue.}
\label{fig:lx}
\end{figure*}

\begin{figure*}
\hbox{
 \includegraphics[width=0.5\textwidth]{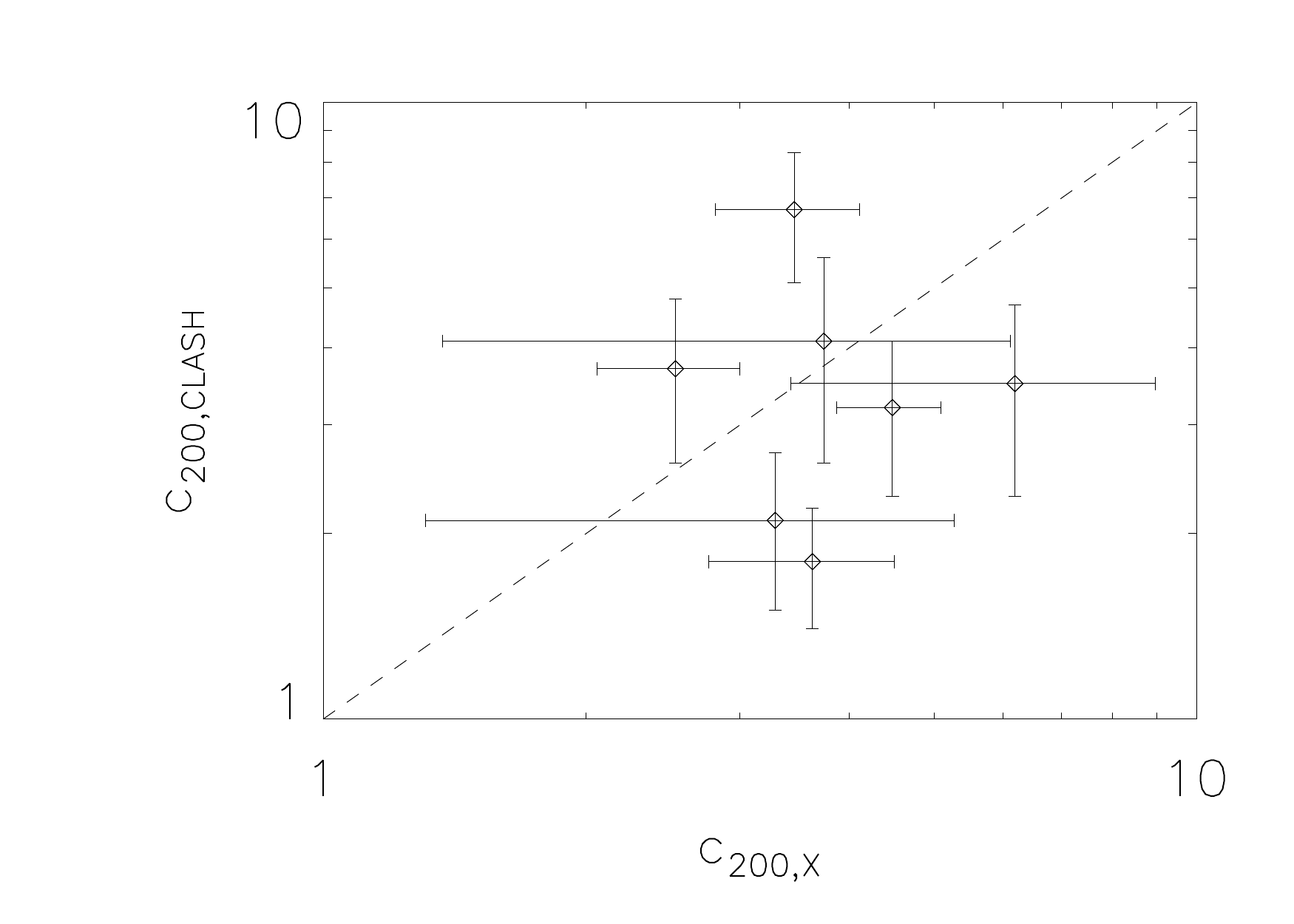}
 \includegraphics[width=0.5\textwidth]{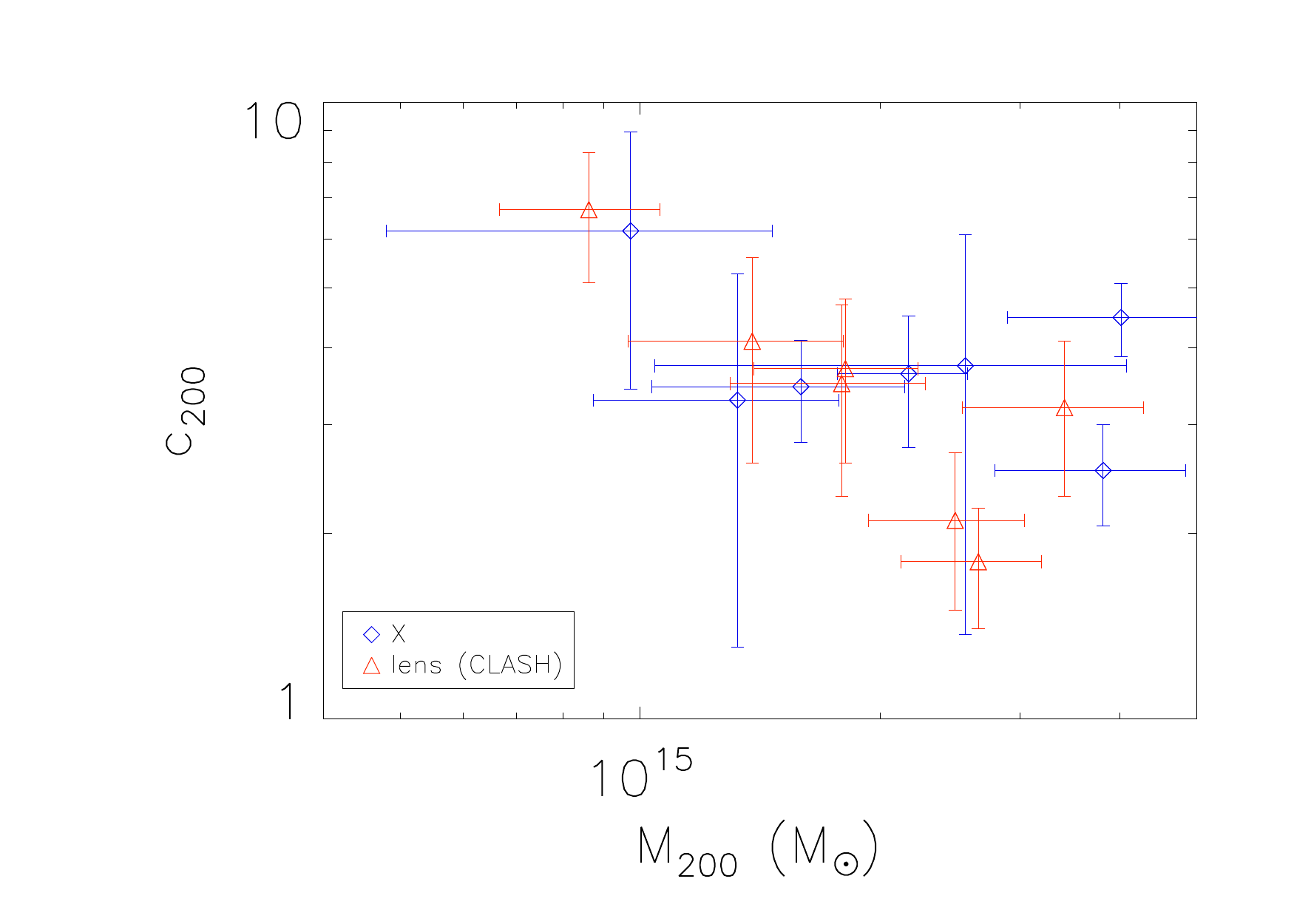}
}
\caption{Comparison between our constraints from X-ray data and CLASH lensing estimates for the 7 objects in common on the mass concentrations ({\it left}) 
and $c-M$ distribution ({\it right}).}
\label{fig:lx_c200}
\end{figure*}

To assess the agreement between the two measurements, we adopt two methods. First, we consider the (natural) logarithm of the mass ratios
\citep{roz+al14,se+et15_comalit_I}. We consider the  \back\  masses. This estimator is not affected by the exchange of numerator and 
denominator. Since quoted errors in compiled catalogs may account for different sources of statistical and systematic uncertainties and 
published uncertainties are unable to account for the actual variance seen in sample pairs, we conservatively perform an unweighted analysis.

The agreement between mass estimates is good, see Fig.~\ref{fig:lx}. For the masses at $R_{200}$, we measure a ratio
$\ln(M_\text{X}/M_\text{lens}) = 0.16 \pm 0.65$, where the first estimate is the median and the second one is the dispersion of the 
distribution of mass ratios. Mass differences are inflated when computed  at $R_{200}$ due to the different volumes. We then also consider 
the masses enclosed within a fixed physical radius, 1~Mpc. We find  $\ln(M_\text{X}/M_\text{lens}) = 0.01 \pm 0.45$.

Seven clusters of our sample are also covered with ground weak lensing studies by the CLASH program. 
\citet{ume+al15} perform a joint shear-and-magnification weak-lensing analysis with additional strong 
lensing constraints of a sub-sample of 16 X-ray regular and 4 high-magnification galaxy clusters in the redshift range $0.19 \la z \la 0.69$. 
For these clusters, we find $\ln(M_\text{X}/M_\text{lens}) = 0.12 \pm 0.58$ at $R_{200}$ and $\ln(M_\text{X}/M_\text{lens}) = -0.32 \pm 0.74$
within 1 Mpc, consistent with the full lensing sample.

Concentrations are consistent too, see Fig.~\ref{fig:lx_c200}. For the seven CLASH clusters, we find 
$\ln(c_\text{200,X}/c_\text{200,lens}) = 0.19 \pm 0.53$.

As a second method, we estimate the mass bias by regressing the hydrostatic against the lensing masses. We follow the approach detailed 
in \citet{se+et15_comalit_I,se+et15_comalit_IV}, which accounts for heteroscedastic errors, time dependence and intrinsic scatter in both the 
independent and the response variable. This accounts for both $M_\text{lens}$ and $M_\text{X}$ being scattered proxy of the true mass. 
We fit the data with the model $M_\text{X,200}= \alpha + \beta \; M_\text{lens,200} +\gamma \; \log (1+z)$.
First, we assume that the mass ratio $M_\text{X,200}/M_\text{lens,200}$ is constant at given redshift ($\beta=1$) and we find $\alpha= 0.08 \pm 0.15$. 
This bias is consistent with what found with the mass-ratio approach described before. 
$\gamma$ is consistent with zero, $\gamma= -0.15 \pm 0.75$, i.e., we cannot detect any redshift dependence in the bias. 
For the scatters, we find $\sigma_{\log(M_\text{lens,200})}=0.11\pm0.07$, and $\sigma_{\log(M_\text{X,200})}=0.04\pm0.04$.
Then, we check the above assumption by letting the slope free. We find $\alpha=0.40\pm0.28$, $\beta=0.74\pm0.20$, $\gamma= -0.32\pm0.73$, 
$\sigma_{\log(M_\text{lens,200})}=0.08\pm0.07$, and $\sigma_{\log(M_\text{X,200})}=0.07\pm0.05$. The slope $\beta$ is fully consistent with one and the 
other parameters are in full agreement with the determination assuming $\beta=1$.

We conclude that the X-ray masses are in very good agreement with the lensing masses, $M_\text{X,200}/M_\text{lens,200}\sim 1$. 
Uncertainties are too large to make statements about deviations from equilibrium or non-thermal contributions that can bias the X-ray masses low 
\citep{se+et15_comalit_I}.
 
\section{The $c_{200} - M_{200}$ relation}
\label{sec:cm}

\begin{figure*}
\hbox{
 \includegraphics[width=0.5\textwidth]{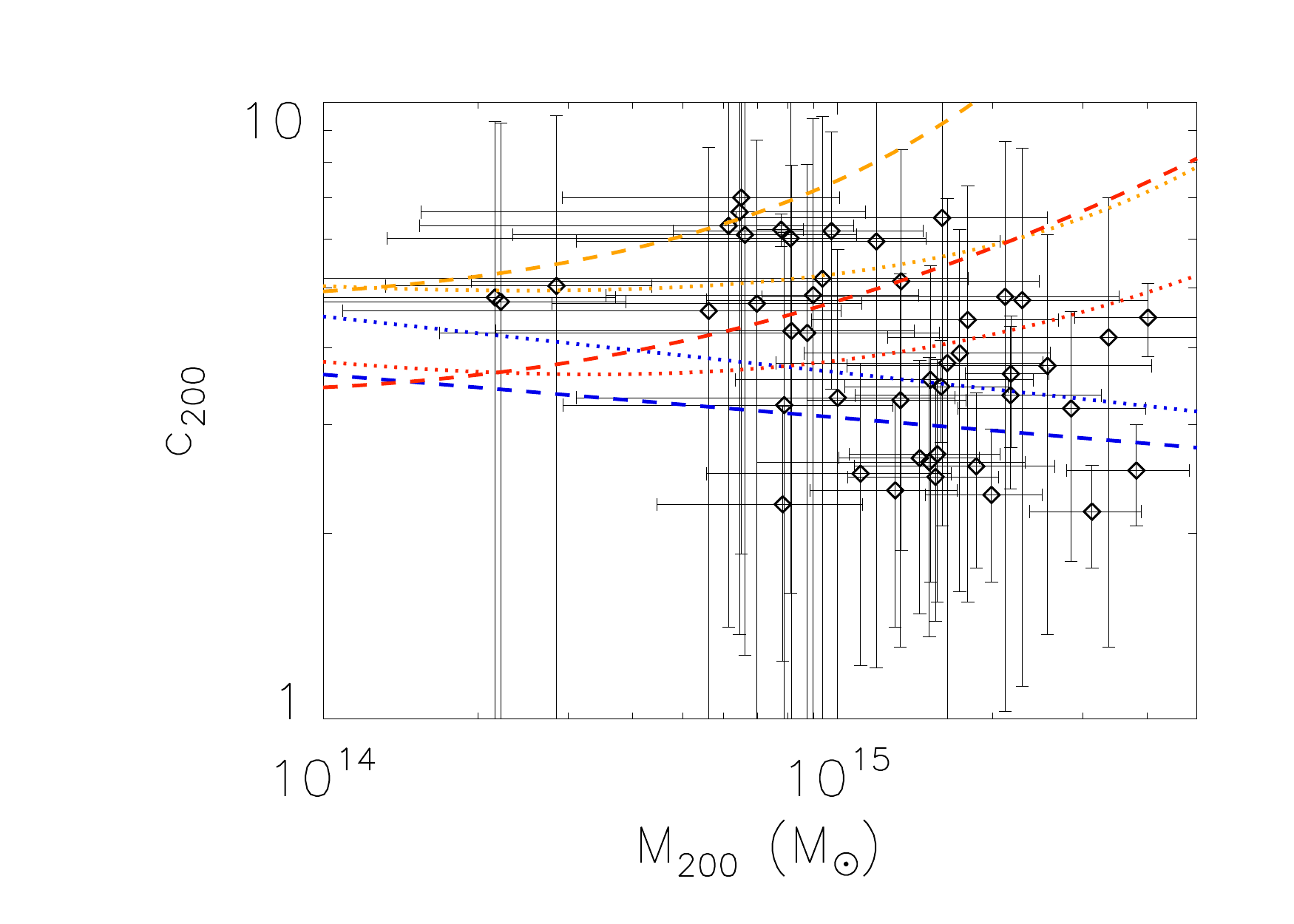}
 \includegraphics[width=0.5\textwidth]{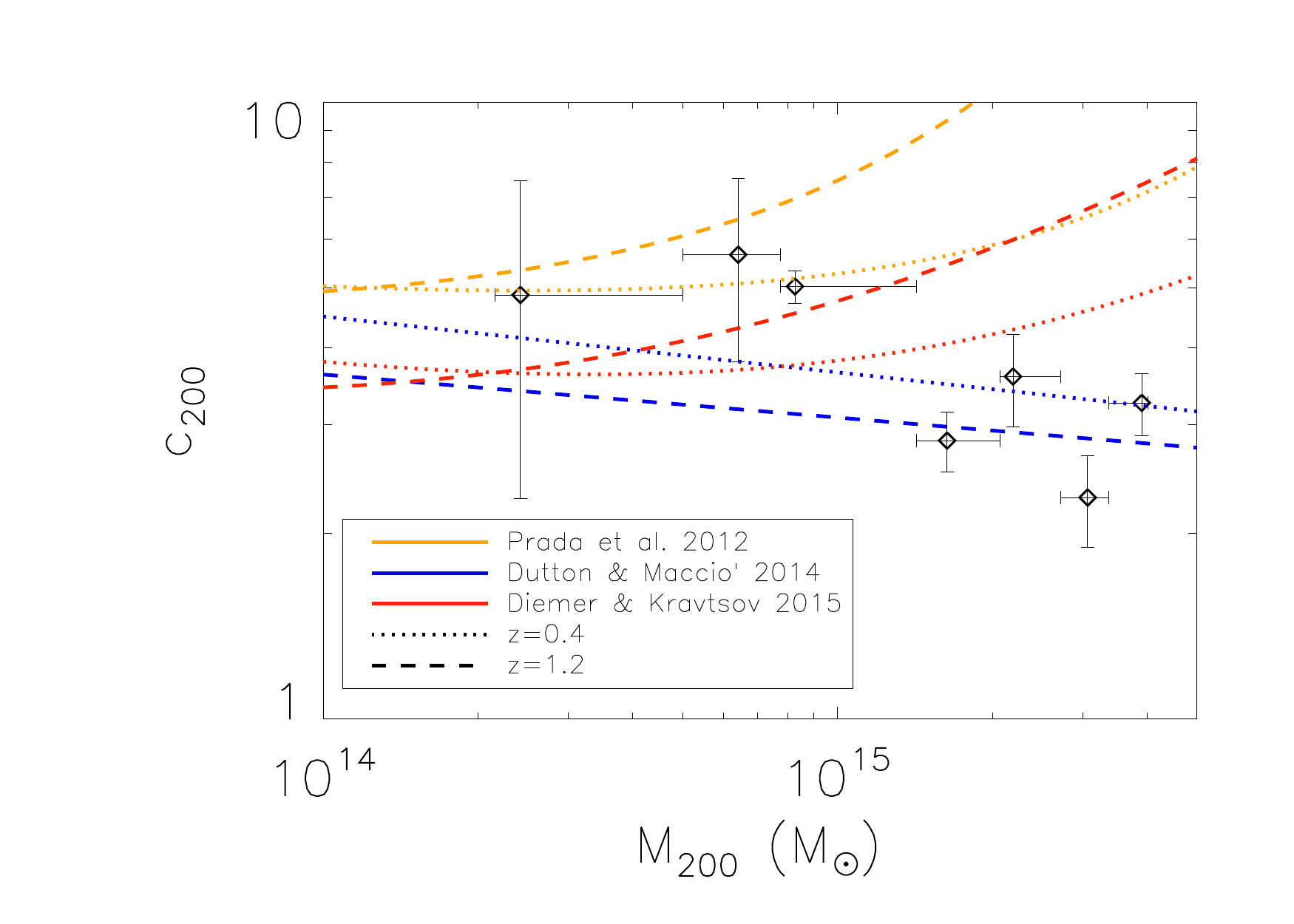}
}
\caption{({\it Left}) Concentration--mass relation obtained for the final cluster sample in the case $\Delta=200$ (black diamonds).
The cluster total masses are obtained following the \back\ described in \citet{ettori+02}.
A NFW profile is adopted to describe the gravitational potential. We overplot the $c_{200}-M_{200}$ relations predicted by P12 
(yellow lines), DM14 (blue lines) and DK15 (red lines). They are calculated for $z=0.4$ (dotted lines) and $z=1.2$ (dashed lines), 
which are the lowest and the highest redshifts in the sample. 
({\it Right}) the same as the left panel but here the sample is divided into 7 mass bins.
For each bin, error-weighted means for concentration and mass are calculated (black diamonds) and the error bars represent the errors on the 
weighted means.} 
\label{fig:cMsim}
\end{figure*}

We present here our results on the $c_{200} - M_{200}$ relation. 
We need to remind that our sample, because of the adopted selection criteria (discussed in Sect.~\ref{sec:dataset}), 
is not statistically complete, but represents well the high mass end of the cluster population, even at high redshift
(see also discussion in Sect.~\ref{sec:complete}).

The concentration-mass relation for the $47$ clusters of our sample is shown in Fig.~\ref{fig:cMsim}. 
The large error bars are due to the uncertainties in determining the observable surface brightness and the spectrum
of each cluster, which are consistently propagated up to the concentration and mass derivation.  

The right panel of Figure \ref{fig:cMsim} is obtained by dividing the sample into seven mass bins and estimating, for each bin, the 
error-weighted mean of the values of the concentration and the error on the mean. This operation is made to enhance the observed signal, 
giving more weight to more precise measurements and to find the mean properties of the sample.

Overall, our data confirm the expected trend of lower concentrations corresponding to higher masses. 
We investigate the distribution of the concentrations for clusters in two mass ranges, respectively below and above the median value of 
$M_{200} = 1.3\times10^{15} M_\odot$. 
The overall distribution is well approximated by a log-normal function with a mean value $<\log c_{200}>$ and a scatter $\sigma$:
\begin{equation}
\small
P(\log c_{200})= \frac{1}{\sigma \sqrt{2\pi}} \exp{\left[-\frac{1}{2}\left(\frac{\log c_{200}-<\log c_{200}>}{\sigma}\right)^2\right]}\,.
\end{equation}
We obtain a mean value for the total concentration distribution of $<\log c_{200}>=0.60$ and a scatter of 
$\sigma(\log c_{200})=0.15$. By considering the two mass ranges, we find a mean of $<\log c_{200}>=0.66$ ($0.54$) and a scatter of
$0.14$ ($0.12$) for the low (high) mass case. 
The central peak is shifted towards the low concentrations in the high mass case, as expected, while we have a slightly larger scatter in 
the low-mass case. 
We also investigate the distribution of the concentrations in two redshift ranges, considering the median redshift of the sample, $z=0.6$ as 
threshold. We find a mean of $<\log c_{200}>=0.55$ ($0.66$) and a scatter of 
$\sigma(\log c_{200})=0.14$ ($0.13$) for the low (high) redshift case, consistent with the above estimates.

In Fig.~\ref{fig:cMsim}, we also compare our data with three recent results from numerical simulations: 
\citet[hereafter DK15]{dk15}, \citet[hereafter DM14]{dm14}, \citet[hereafter P12]{prada+12}. 
The range of the predicted results is delimited by a dotted line, corresponding to the lowest redshift in the sample ($z=0.4$), 
and a dashed line, corresponding to the highest one ($z=1.2$).
The comparison with these theoretical works is carried out using the public code {\it Colossus} 
provided by DK15\footnote{\url{www.benediktdiemer.com/code}}.
It is a versatile code that implements a collection of models for the c--M relation, including the ones of our interest, 
allowing the choice of a set of cosmological parameters and the conversion among different mass definitions. 
It turns out to be very useful for our purpose, in order to homogenize the results presented in the original papers 
to our cosmological model of reference and to masses defined at $\Delta=200$ with respect to the critical 
density of the Universe, as in our analysis.

However, it must be noted that we investigate a mass range that might exceed slightly those probed by numerical simulations, in particular at $z\sim1$. 
In fact, there are no numerical predictions for the behaviour of the $c-M$ relation for masses larger than $10^{15} M_\odot$ in the range of redshifts considered in our work.
To make the comparison with our results, we proceed by using the numerical predictions as extrapolated from the available datasets 
\footnote {in the case of the adopted code  {\it Colossus}, see the description of the models implemented at \url{bdiemer.bitbucket.org/halo\_concentration.html}.}.

In order to quantify the deviations from numerical predictions, we use the following $\chi^2$ estimator:
\begin{equation}
\label{eq:obs-sim}
 \chi^2 = \sum_i \frac{(\log c_\text{obs,i}(M,z)-\log c_\text{sim,i}(M,z))^2}{\epsilon_{\log c_\text{obs,i}}^2 + \sigma_{\log c_\text{sim}}^2},
\end{equation}
where the sum is done over the 47 clusters of our sample; $c_\text{obs}$ and $\epsilon_{c_\text{obs}}$ are the estimates of concentrations and
the corresponding errors, respectively, listed in Table~\ref{tab:cm} 
(we omit the label \textquotedblleft 200\textquotedblright \ to simplify the notation); $c_\text{sim}$ are the values derived from the models
for fixed mass and redshift; $\sigma_{\log c_\text{sim}}$ is the intrinsic scatter on the simulated concentrations, assumed to be equal to 
0.11 (e.g. DM14). 
We obtain a $\chi^2$ of 272.4, 26.3 and 69.4 when the models by P12, DM14 and DK15, respectively, are considered. 
A random variable from the $\chi^2$ distribution in equation~(\ref{eq:obs-sim}) with 47 degrees of freedom has a probability of 100, 0.6 and 98 per cent to be lower than 
the measured values, respectively, indicating a tension with the P12 and DK15 models in the mass ($10^{14}-4\times 10^{15} M_{\odot}$) 
and redshift ($0.4-1.2$) ranges investigated in the present analysis.

It is clear from the right panel of Figure~\ref{fig:cMsim} that our results show the lowest concentrations for the highest masses, and 
are not compatible with an upturn at the high masses.
This is indeed expected for a sample of relaxed clusters only \citep[see e.g.][]{ludlow+12,correa+15}. We note that the models considered here characterise different halo samples: P12 and DK15 include all the halos, regardless their degree of virialization, whereas DM14 exclude the unrelaxed halos. Even though the selection is different, we remark that we are considering objects that show no major mergers and are closer to the selection for relaxed halos applied in numerical simulations.
Moreover, it is worth noticing that the concentrations calculated in P12 are derived from the circular ratio $V_\text{max}/V_{200}$, 
rather than from a direct fit to the mass profile and that the halos are binned according to their maximum circular velocity, rather than in 
mass. As pointed out in \citet{mr13}, such methodological differences lead to large discrepancies both in the amplitude and in the shape of the $c-M$ relation, 
especially on the scales of galaxy clusters, making the comparison with the predictions in P12 not straightforward.

\subsection{Evolution with redshift}
\label{sec:cmz}

\begin{figure*}
\centering
\includegraphics[width=0.33\textwidth]{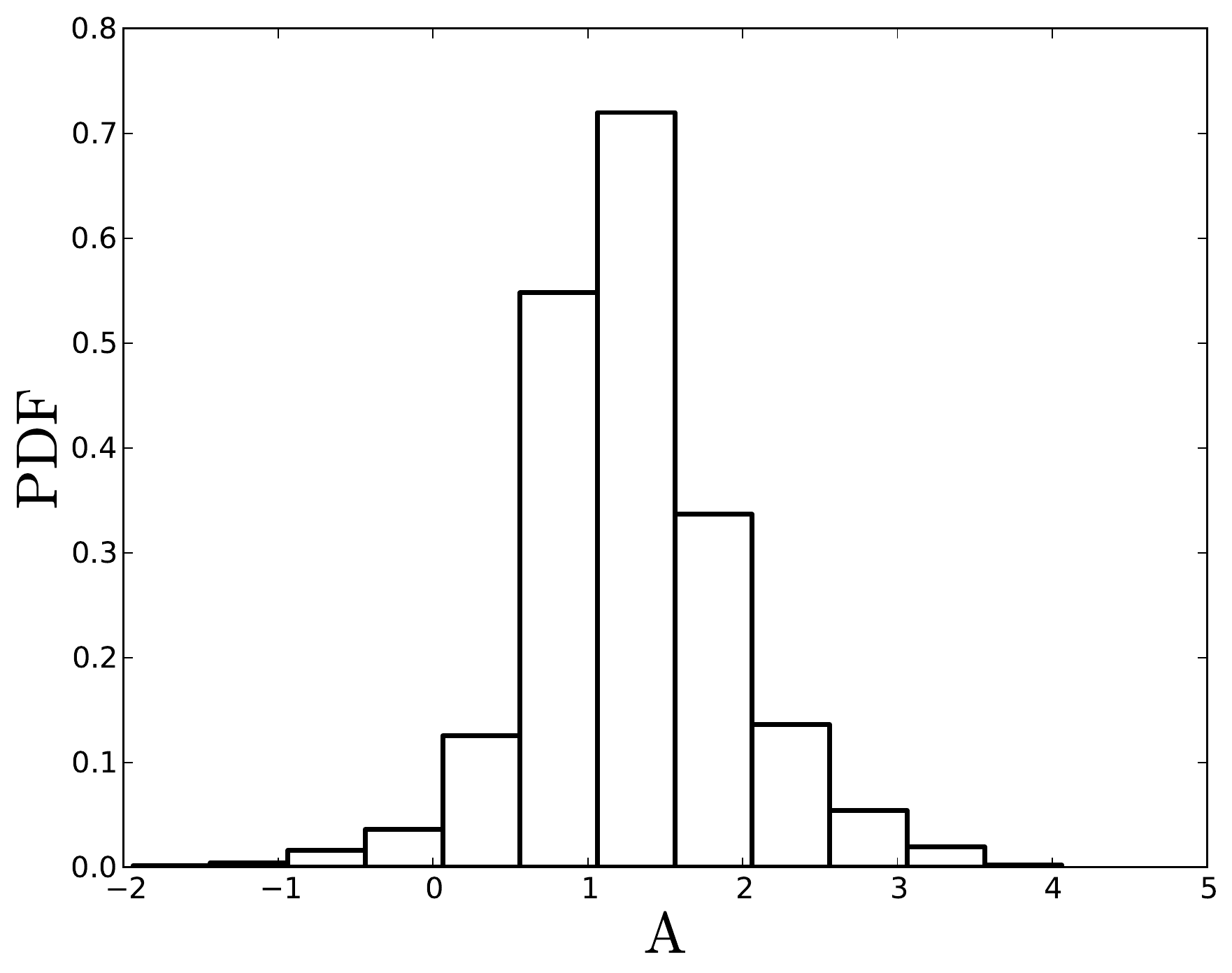}
\includegraphics[width=0.33\textwidth]{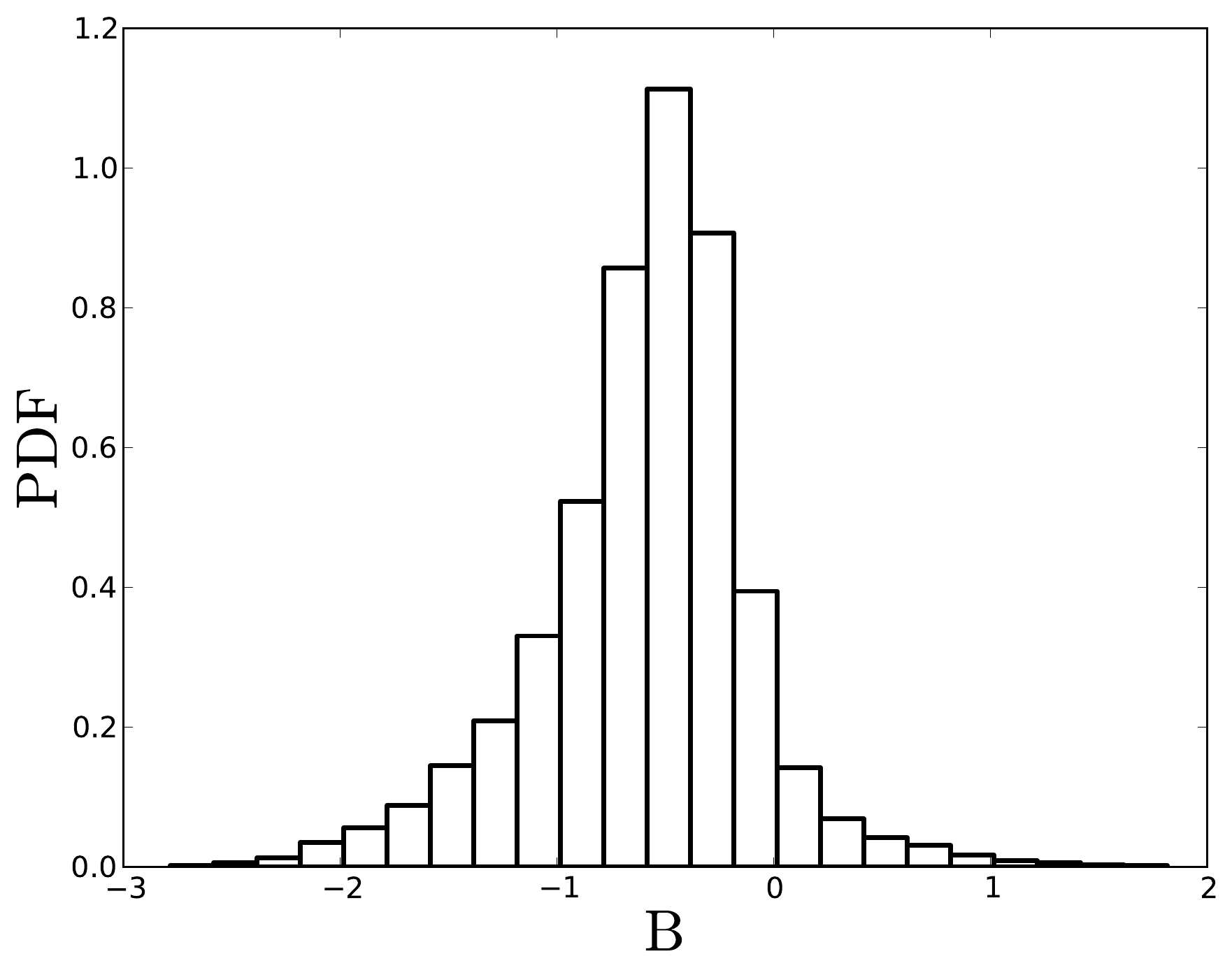}
\includegraphics[width=0.33\textwidth]{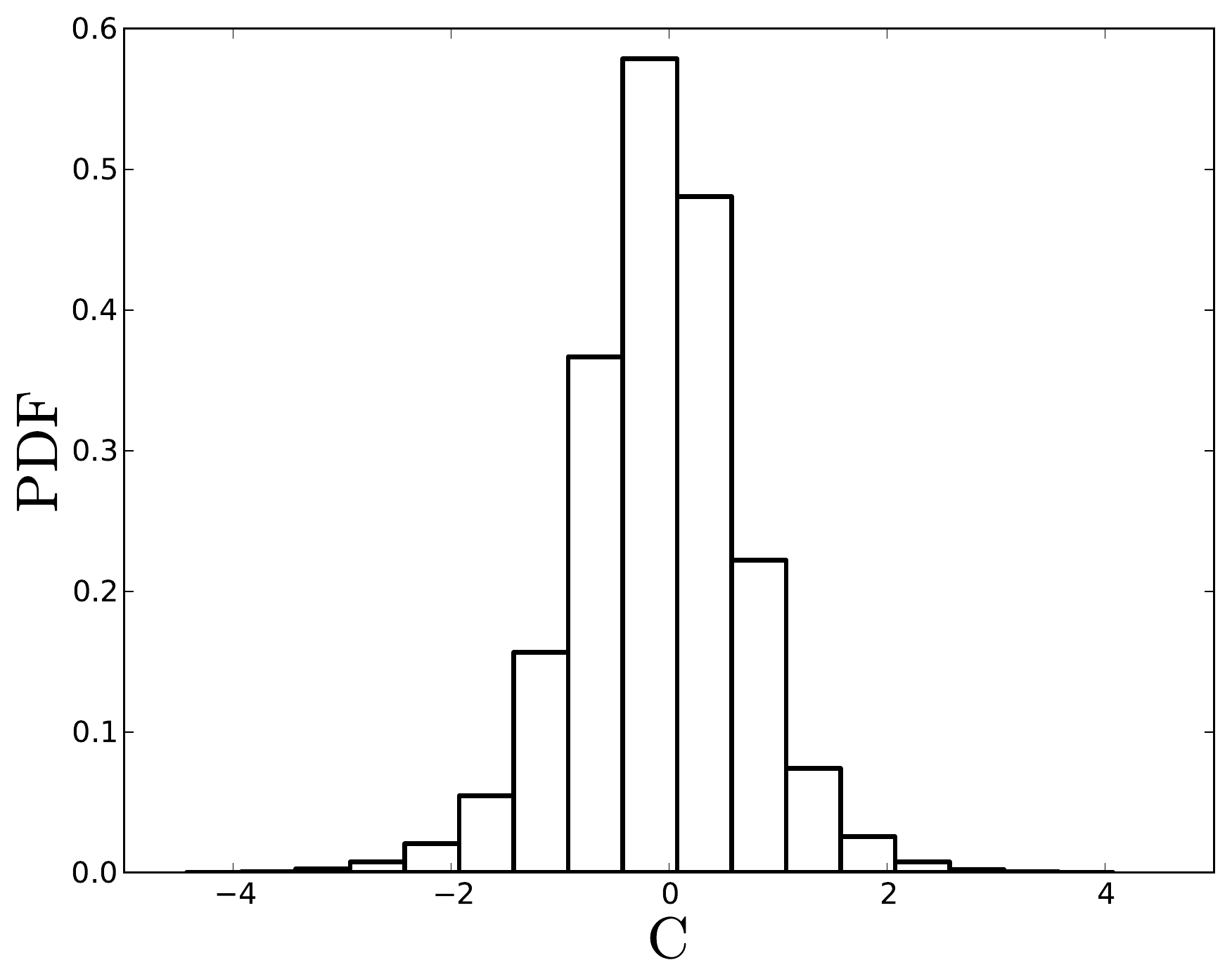}
\includegraphics[width=0.33\textwidth]{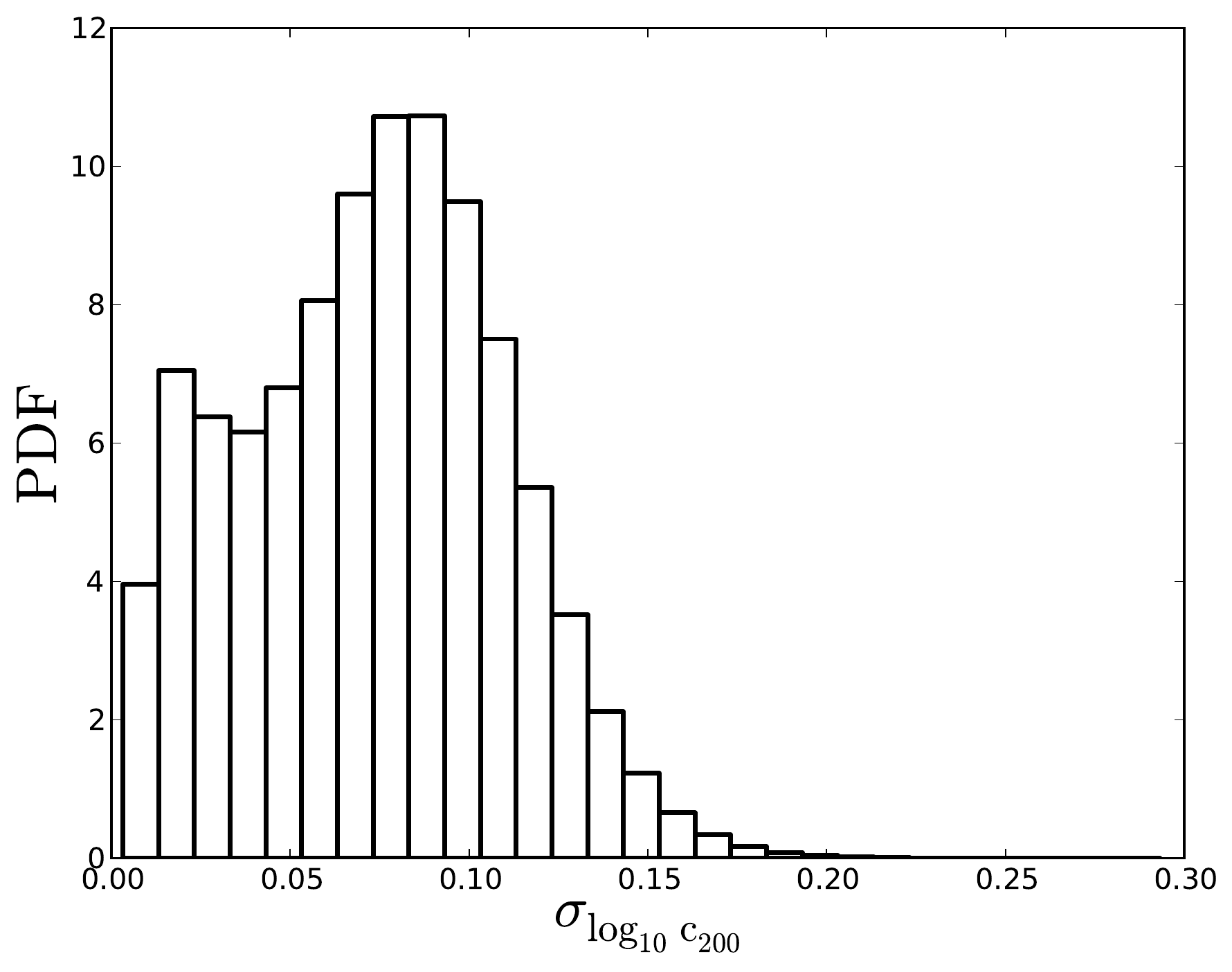}
\includegraphics[width=0.33\textwidth]{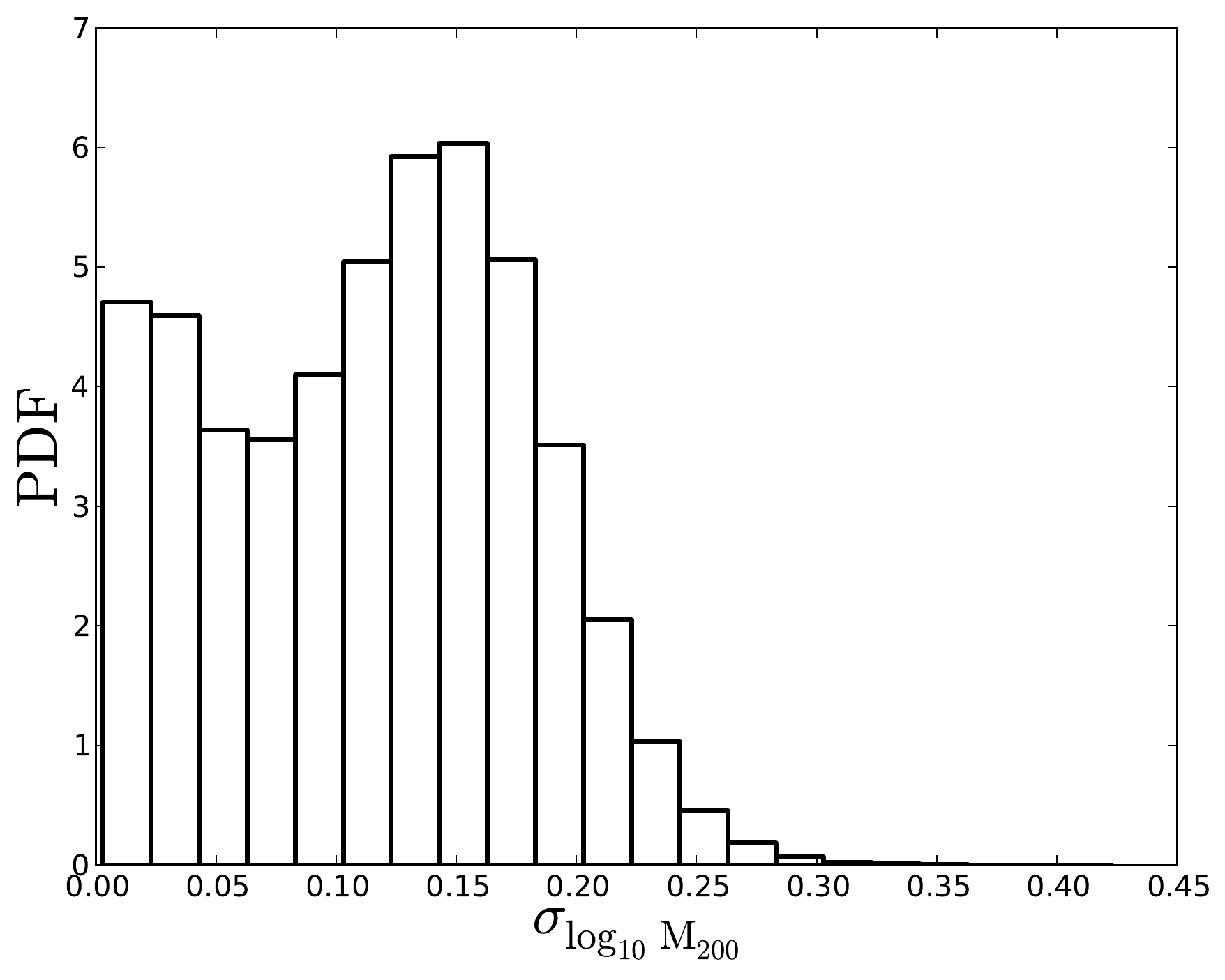}
\caption{Probability distributions of the best-fit parameters of the c--M--z relation eq.(\ref{func:fit}) obtained with LIRA,
where the covariance between mass and concentration is taken into account.}
\label{fig:1Dprob} 
\end{figure*}

\begin{figure}
\centering
\includegraphics[width=0.4\textwidth]{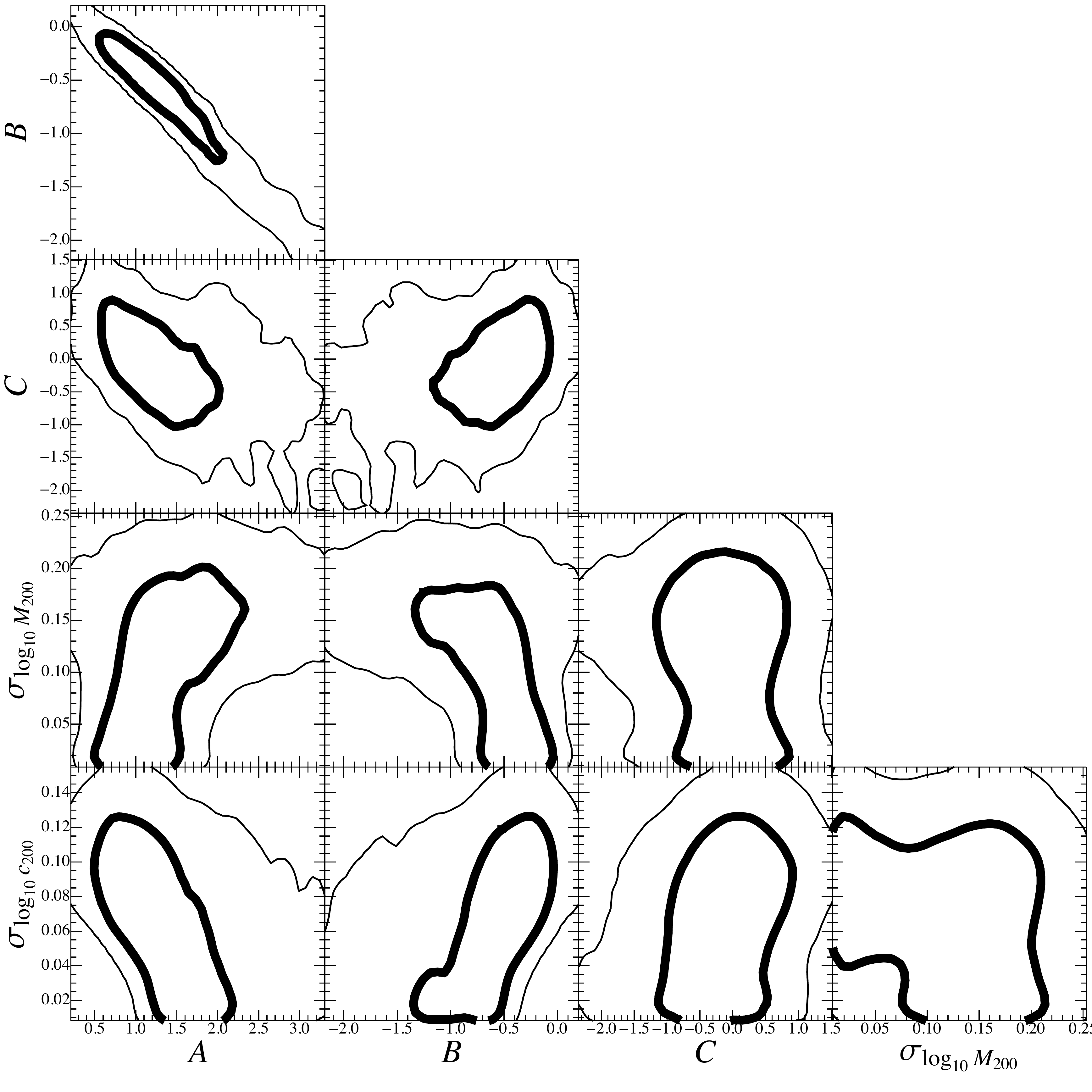}
\caption{Probability distributions of the best-fit parameters of the c--M--z relation eq.(\ref{func:fit}) obtained with LIRA, 
where the covariance between mass and concentration is taken into account. 
The thick (thin) lines include the 1-(2-)$\sigma$ confidence region in two dimensions, here defined as the region within which the value of 
the probability is larger than $\exp[-2.3/2]$ ($\exp[-6.17/2]$) of the maximum.}
\label{fig:2Dcontours} 
\end{figure}

\begin{figure}
\includegraphics[width=0.5\textwidth]{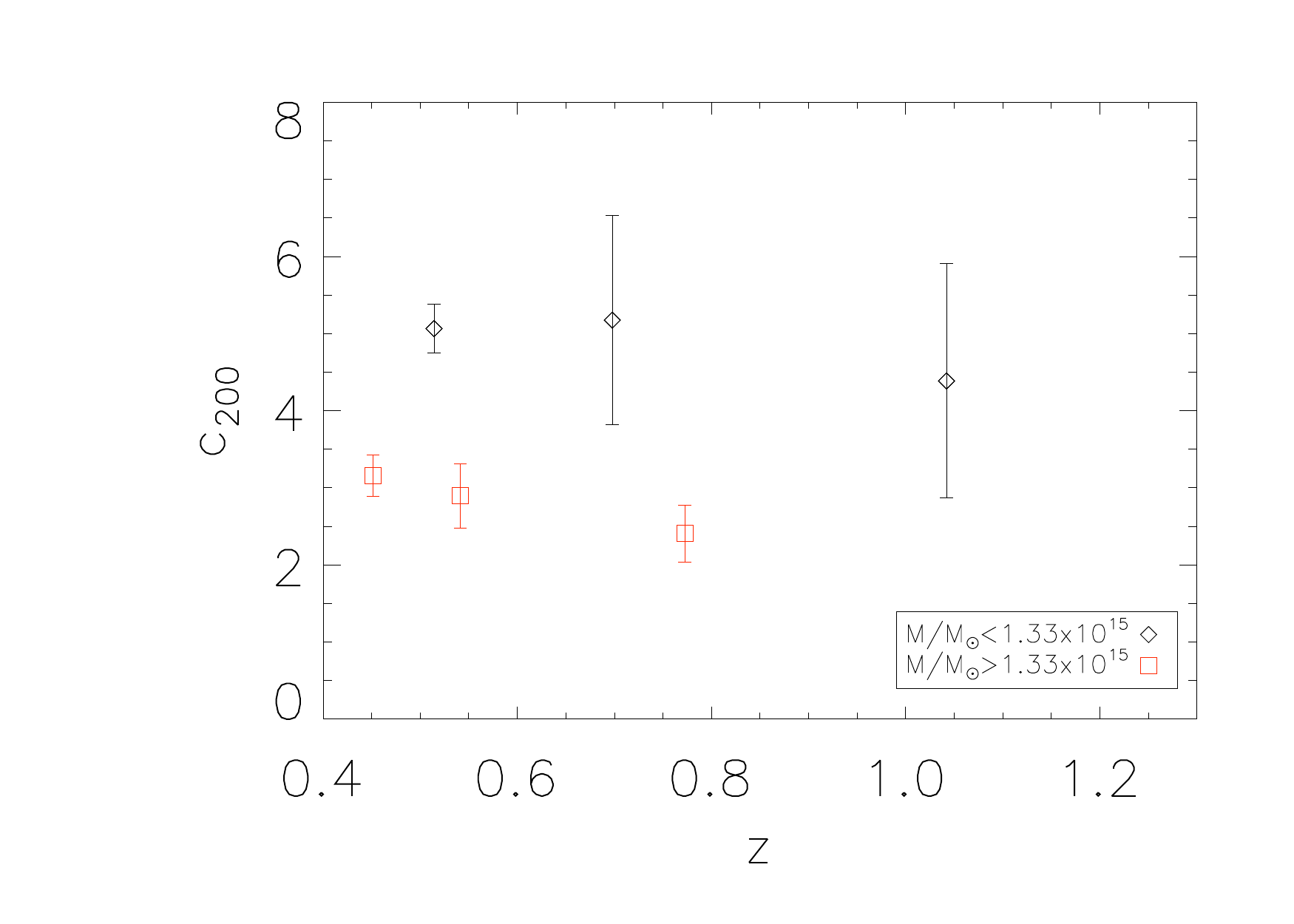}  
\caption{Concentration-redshift relation calculated in two mass ranges: $M\leq1.33\times10^{15} M_\odot$ (black) and $M<1.33\times10^{15} M_\odot$
(red). For each mass range, the points are the error-weighted means of the concentrations and the error bars are the errors on the means, for 
three redshift bins. The sample is approximately evenly divided in each bin and we show the median redshift for simplicity.
}
\label{fig:cz}
\end{figure}

\begin{figure}
\centering
\includegraphics[width=0.4\textwidth]{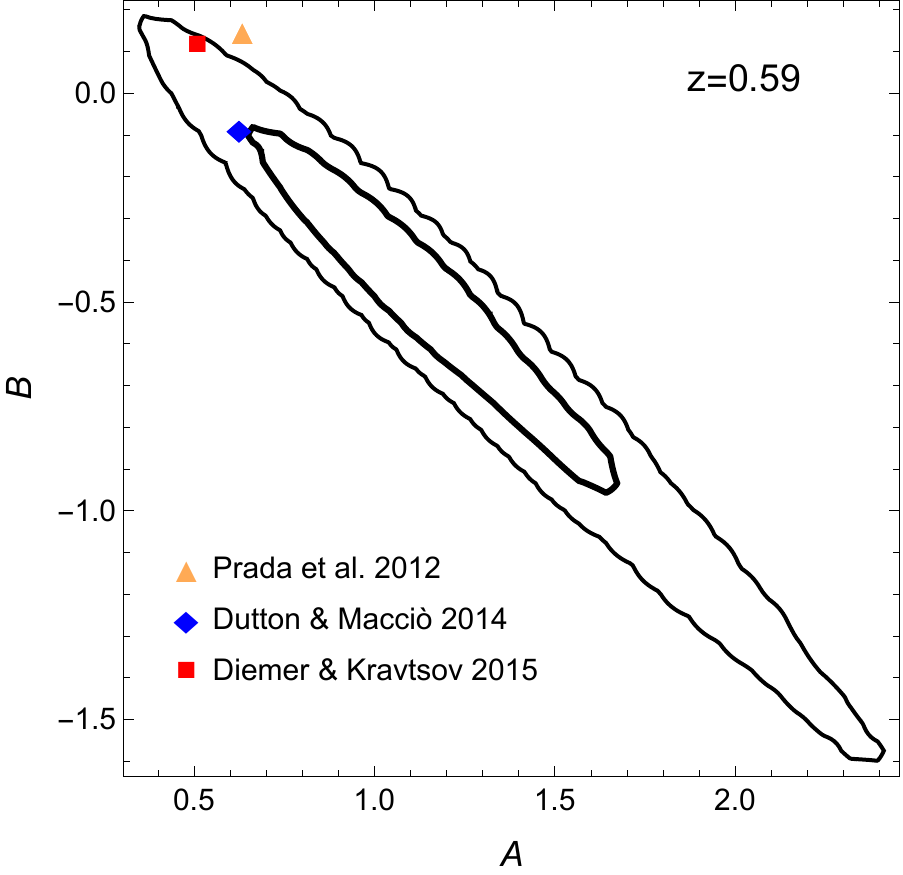}  
\includegraphics[width=0.4\textwidth]{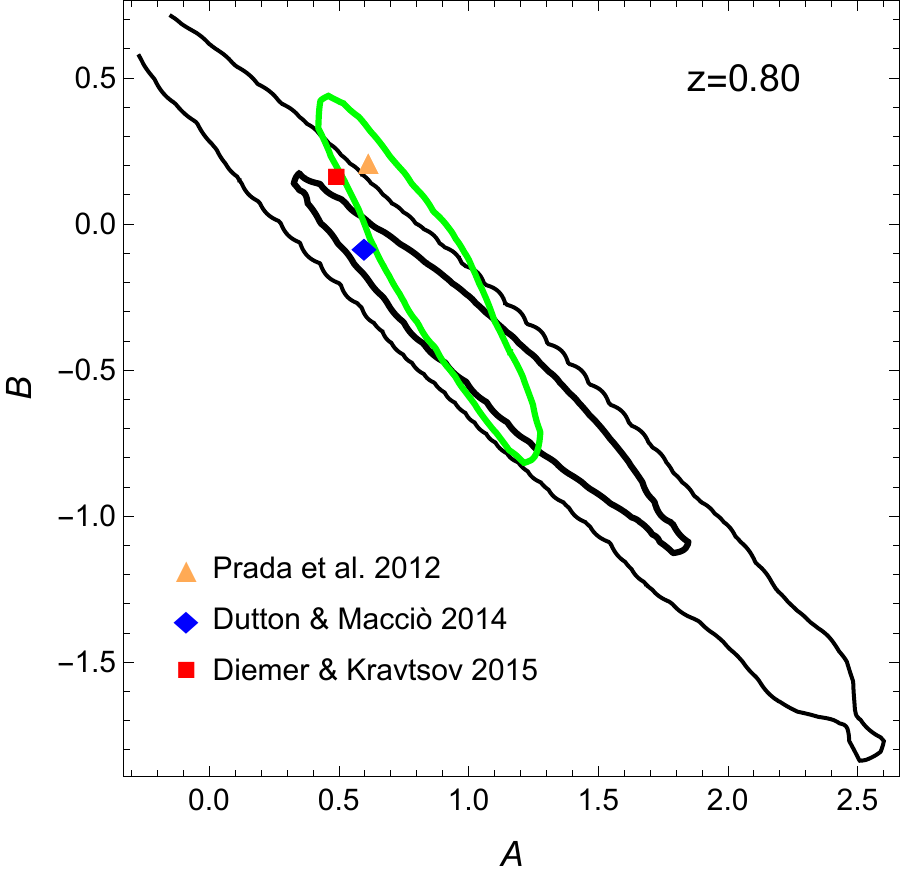}  
\caption{
Probability distributions of the A and B parameters of the c--M relation eq.(\ref{func:fit}) calculated with LIRA, 
for the full sample ({\it Top}) and for the subsample of clusters at $z \ge 0.7$ ({\it Bottom}). The relations are normalised 
at the median redshift of the sample considered ($0.59$ and $0.80$, respectively).
The confidence regions are defined as in Fig.~\ref{fig:2Dcontours}. 
The coloured symbols show the estimates of the parameters from simulations by P12, DM14 and DK15 evaluated at the quoted redshift.
The green contour shows the constraints from \citet{sc13} at $1 \sigma$.
}
\label{fig:2DcontoursAB} 
\end{figure}

With the aim of investigating the dependence of the cluster concentrations on mass and redshift, we consider the three-parameter functional 
form, $c = c_0 M^B (1+z)^C$, and we linearly fit our data to the logarithmic form of this function:
\begin {equation}
\log c_{200}=A+B\,\log \left(\frac{M_{200}}{10^{14}\,M_\odot} \right)+C\,\log(1+z) \pm \sigma_{\log c_{200}} \,.
\label{func:fit}
\end{equation}
We use the Bayesian linear regression method implemented in the R package \texttt{LIRA} by \citet{sereno16}.
We assume a uniform prior for the intercept $A$ and a Student's-t prior for both the mass slope $B$ and the slope of the time evolution $C$. 
For the intrinsic scatter, we assume that $1/\sigma_{\log c_{200}}^2$ follows a gamma distribution.
We obtain the following best-fit parameters: $A=1.15 \pm 0.29$ and $B=-0.50 \pm 0.20$, $C=0.12 \pm 0.61$ and an intrinsic scatter 
$\sigma_{\log c_{200}}=0.06\pm0.04$.
This value is lower than the estimates presented in Sect.~\ref{sec:cm} since here we are correcting for the intrinsic scatter in the hydrostatic masses.
The additional correction for this intrinsic scatter of the mass distribution steepens the relation. On the other hand, by taking into account
the covariance between mass and concentration, we find a flatter relation, as already pointed out from previous work \citep[e.g.][]{sc13}.

These values are fully consistent, within the estimated errors, with the IDL routine \texttt{MLINMIX\_ERR} by \citet{kelly07}, which also 
employs a Bayesian method and with the \texttt{MPFIT} routine in IDL \citep{williams+10,markwardt09} that looks for the minimum of the 
$\chi^2$ distribution by taking into account the errors on both the variables. 
We quote the best-fit values in Table \ref{tab:fit_param}. The probability distributions of the best-fit values obtained with LIRA are shown
Fig.~\ref{fig:1Dprob}, while the two-dimensional 1-(2-)$\sigma$ confidence regions are shown in Fig.~\ref{fig:2Dcontours}.

\begin{table*}
\caption{Best-fit values of the c--M--z relation.} 
\label{tab:fit_param}
\vspace*{-0.2cm}
\begin{center}
\begin{tabular}{c c c c c c}
\hline\hline
Method & $A$ & $B$ & $C$ & $\sigma_{\log_{10}c_{200}}$ & $\sigma_{\log_{10}M_{200}/10^{14}}$  \\ 
\hline
LIRA & $1.15 \pm 0.29$ & $-0.50 \pm 0.20$ & $0.12 \pm 0.61$ & $0.06 \pm 0.04$ & $...$ \\
LIRA (covxy) & $1.23 \pm 0.55$ & $-0.54 \pm 0.41$ & $-0.08 \pm 0.69$ & $0.07 \pm 0.04$ & $0.12 \pm 0.07$ \\
LIRA ($C=0$) & $1.19 \pm 0.24$ & $-0.51 \pm 0.20$ & $0$ & $0.06 \pm 0.04$ & $...$ \\
LIRA ($B=-0.1$) & $0.61 \pm 0.12$ & $-0.10$ & $0.38 \pm 0.64$ & $0.10 \pm 0.02$ & $...$ \\
MLINMIX\_ERR & $1.07 \pm 0.37$ & $-0.42 \pm 0.21$ & $-0.02 \pm 0.97$ & $0.09 \pm 0.03$ & $...$ \\
MPFIT & $1.34 \pm 0.15$ & $-0.53 \pm 0.07$ & $-0.57 \pm 0.65$ & $...$ & $...$ \\
\hline
\end{tabular}
\end{center}
\tablefoot{The best fit parameters refer to equation~(\ref{func:fit}) and are obtained using: 
two Bayesian multiple linear regression methods, \texttt{LIRA} and \texttt{MLINMIX\_ERR}; the linear least square fitting \texttt{MPFITFUN}.
All the methods account for heteroscedastic errors in both the independent and the dependent variables.
} 
\end{table*}

We measure a normalisation $A\approx 1$ and a mass slope $B\approx-0.5$, lower than the value predicted by numerical simulations 
($-0.1$). By fixing the parameter $B$ to $-0.1$, we find $A=0.61 \pm 0.12$, $C=0.38 \pm 0.64$ and 
$\sigma_{\log c_{200}}=0.10 \pm 0.02$.

With the Bayesian methods we measure a typical error that is larger by a factor of $2$ in normalisation and by a factor of $2.5$ 
in the mass slope, with respect to the corresponding values obtained through the covariance matrix of the \texttt{MPFIT} method. 
All the methods estimate large errors in the redshift dependence and the best-fit values of the redshift slope are consistent with zero
(at $1\sigma$ level).

The concentration-redshift relation is shown in Fig.~\ref{fig:cz} for clusters in two mass ranges, considering the median mass 
$1.33\times10^{15} M_\odot$ as threshold. For each mass range, the sample is divided into three redshift bins, chosen to have approximately an 
equal number of clusters in each bin:
[0.426-0.583], [0.600-0.734], [0.810-1.235] for the low-mass case, [0.412-0.494], [0.503-0.591], [0.700-0.888] for the high-mass case.
For each bin we calculate the error-weighted means of the concentrations and the errors on the means, obtaining:
$5.06\pm0.31$, $5.18\pm1.36$, $4.39\pm1.52$ for the low-mass case, $3.16\pm0.27$, $2.90\pm0.42$, $2.41\pm0.37$ for the high-mass case.
At a fixed mass range, the concentration slightly decreases with redshift, as expected by the 
fact that the cluster's concentration is determined by the density of the Universe at the assembly redshift. 

Finally, we test the c--M relation in the high redshift regime against the different theoretical models.
We use models by P12, DM14 and DK15 to obtain predictions on the measurements of the normalisation and slope of the c--M relation at the median
redshift of our sample, in the mass range ($10^{14}-4\times 10^{15} M_{\odot}$) investigated in the present analysis (we consider 50 log-mass 
constant points for the fit). As we show in Fig.~\ref{fig:2DcontoursAB}, the predictions from numerical simulations are well in agreement 
with our constraints, with values from DM14 model that are consistent at $1 \sigma$ level, and with larger deviations 
(but still close to the $\sim 2 \sigma$ confidence level) associated to the P12 and DK15 expectations. 

In particular, once we consider only the 18 clusters of our sample with $z \ge 0.7$ and we re-calculate the $\chi^2$ estimator 
in equation~(\ref{eq:obs-sim}), we obtain 62.3, 6.1 and 17.7 when the models by P12, DM14 and DK15, respectively, are considered. 
This means that a random variable from the $\chi^2$ distribution with 18 degrees of freedom has a probability of 99.9, 0.4 and 52.5 per cent 
to be lower than the measured values, respectively, indicating that 
only the P12 model deviates more significantly from our estimates in the ($0.7-1.2$) redshift range.

\section{Sample properties}
\label{sec:properties}

As discussed in Sect.~\ref{sec:dataset} and \ref{sec:analysis}, a cluster at $z>0.4$, and with exposures available in the \chandra\ archive, 
is included in our sample if 1) it is observed with sufficient X-ray counts statistics to get a temperature profile with at least three radial bins; 
2) to a visual inspection of the X-ray maps, it appears to have a regular morphology, so that we can consider it to be close to the hydrostatic equilibrium. 
Thus, we exclude the objects with a strongly elongated shape or the ones containing major substructures. Although we do not use any quantitative criterion 
for this selection, we provide here a morphological analysis in order to present the statistical properties of the sample and to allow a comparison with other 
X-ray samples. We analyse the morphology of each cluster according to the following two indicators: the X-ray brightness concentration
parameter, $c_{SB}$, defined as the ratio between the surface brightness, $S_b$, within a circular aperture of radius 100 kpc and the surface 
brightness enclosed within a circular aperture of 500 kpc:

\begin{equation}
c_{SB} = \frac{S_b (r<100\,kpc)}{S_b (r<500\,kpc)} \,;
\end{equation} 

the centroid shift, $w$, calculated as the standard deviation of the projected separation between the X-ray peak and the centroids estimated 
within circular apertures of increasing radius, from 25 kpc to $R_{ap}$ = 500 kpc, with steps of 5\%:

\begin{equation}
w = \left[ \frac{1}{N-1} \sum_i (\Delta_i - <\Delta>)^2 \right]^{1/2} \frac{1}{R_{ap}} \,,
\end{equation}
where $\Delta_i$ is the distance between the X-ray peak and the centroid of the i-th aperture. 

\begin{figure}
\includegraphics[width=0.5\textwidth]{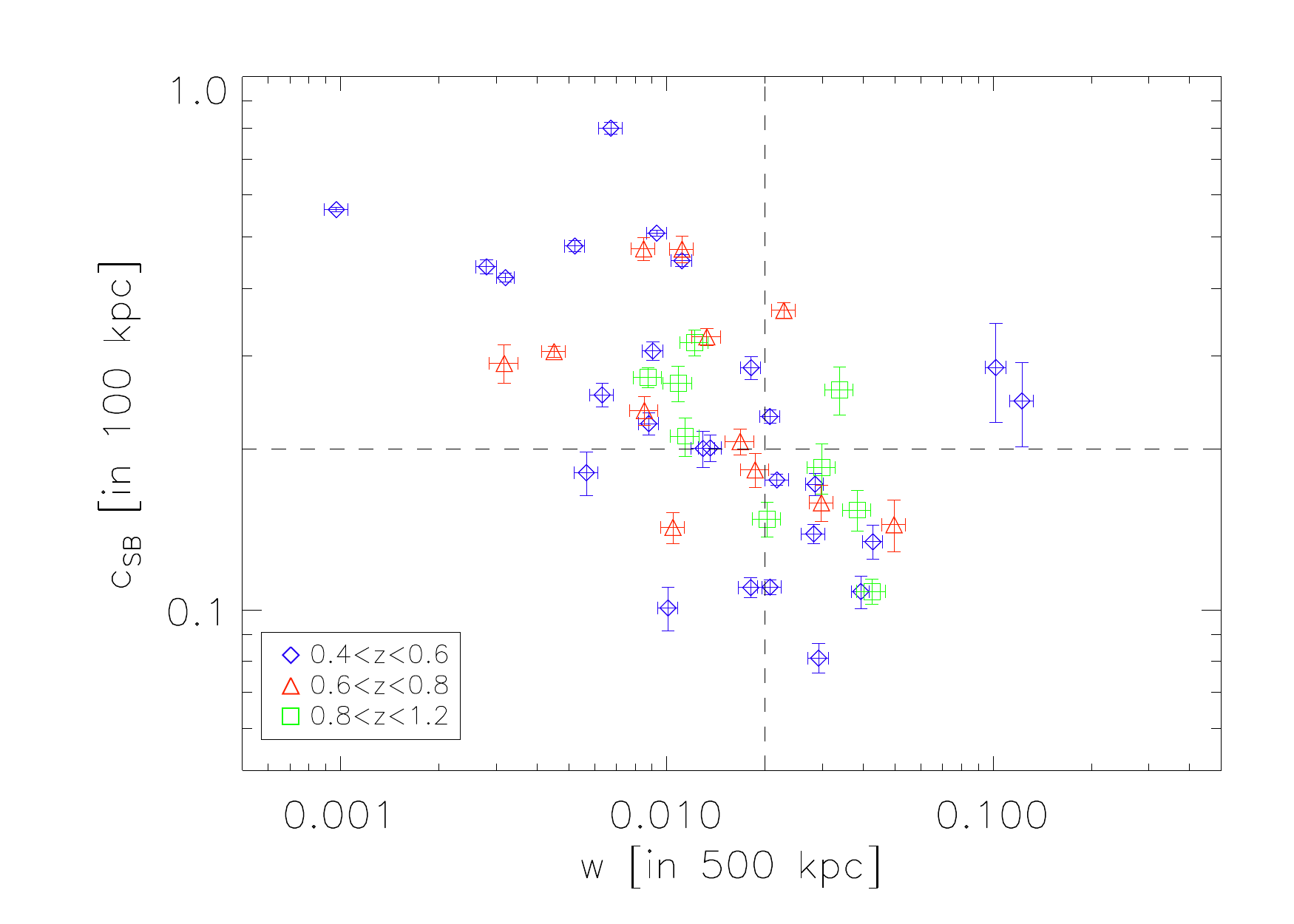}
\caption{Relation between the X-ray brightness concentration and the centroid shift. Dashed lines trace the thresholds indicated by 
\citet{cassano+10} to define relaxed and disturbed clusters (see text). Different symbols and colours are used for clusters in different 
redshift intervals.}
\label{fig:morphology}
\end{figure}

Figure~\ref{fig:morphology} shows that the X-ray concentration is anti-correlated with the centroid shift, qualitatively following the relation
found by \citet{cassano+10}. According to their results, clusters with $c_{SB}>0.2$ and $w<0.012$ are classified as ``relaxed'' (upper left
quadrant in Fig.~\ref{fig:morphology}), while those with $c_{SB}<0.2$ and $w>0.012$, about 1/3 in our sample, are classified as ``disturbed'' 
(lower right quadrant). We note that the relative composition of relaxed/disturbed clusters changes with the redshift, with the 50\% of 
the clusters observed at $z>0.8$ being disturbed. 

\begin{figure}
\begin{tabular}{c}
\includegraphics[width=0.5\textwidth]{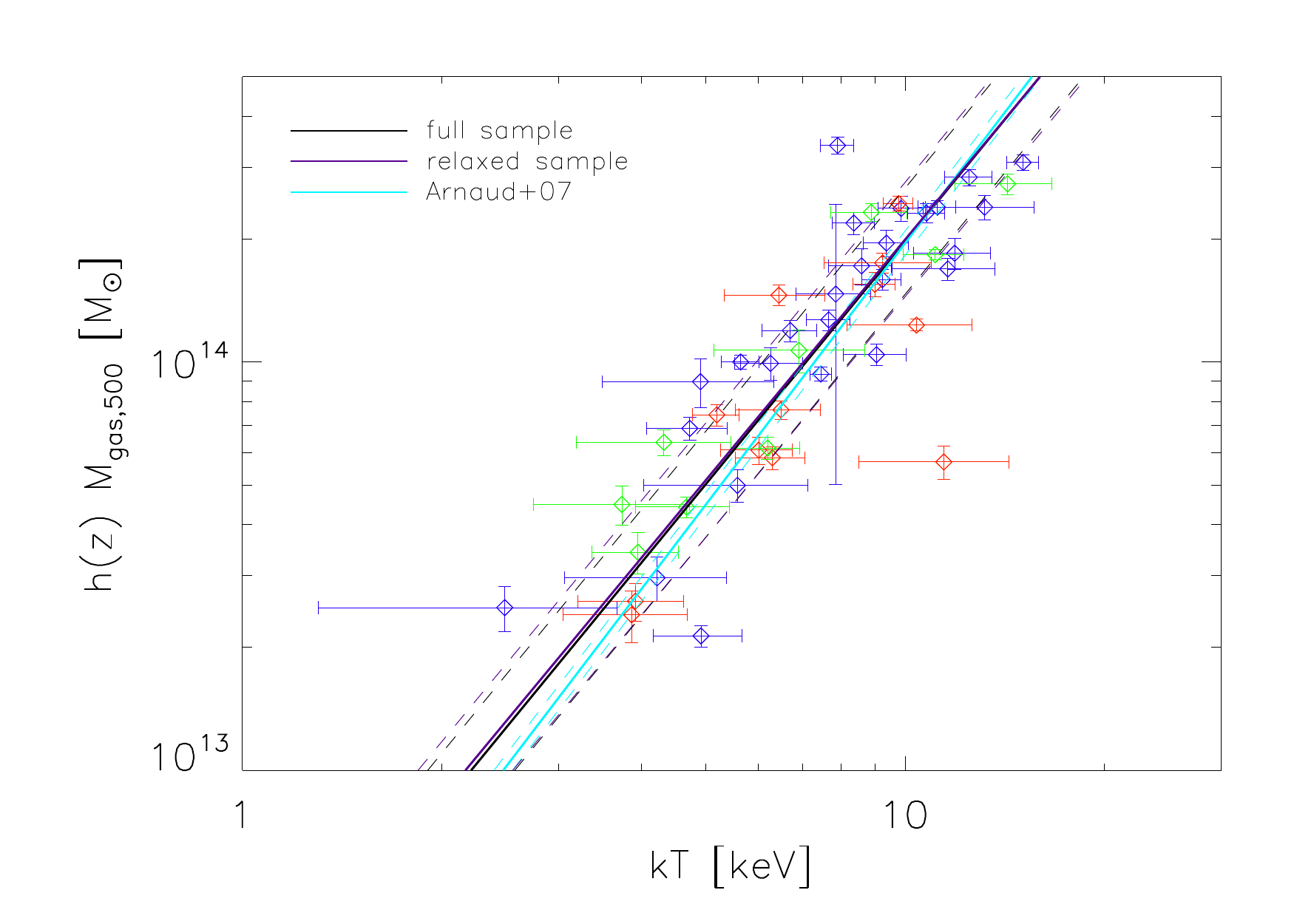} \\
\includegraphics[width=0.5\textwidth]{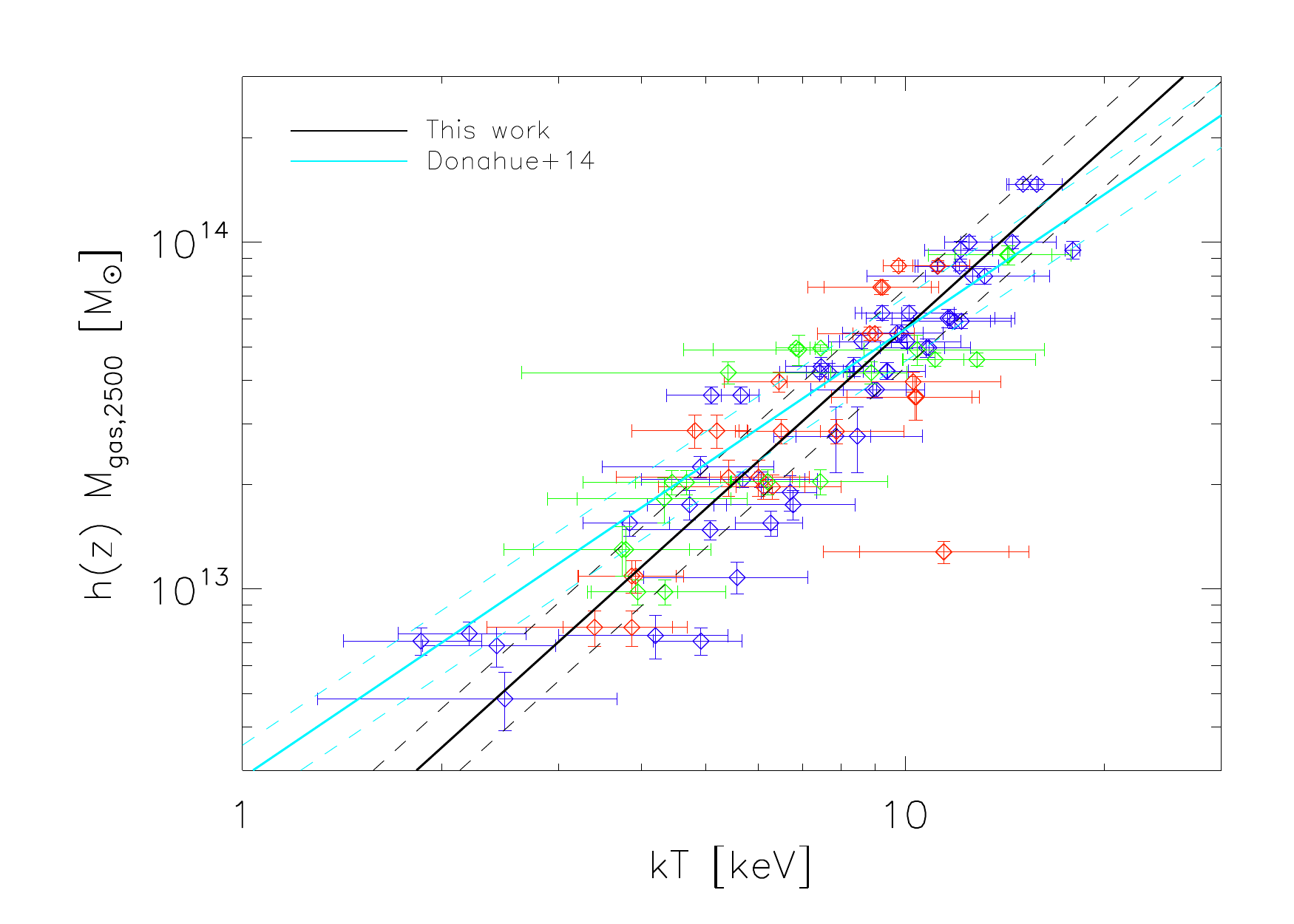} \\
\end{tabular}
\caption{(Top panel) Relation between the gas mass within $R_{500}$ and temperature. 
The black and purple curves show the best-fit relation and its intrinsic scatter obtained for our full and relaxed sample, respectively. 
The cyan curves represents the relation of \citet{arnaud+07}.
(Bottom panel) Relation between the gas mass within $R_{2500}$ and temperature. 
The black curves show the best-fit relation and its intrinsic scatter. 
The cyan curves represents the relation obtained for the CLASH sample.
Coloured symbols as in fig. \ref{fig:morphology}.}
\label{fig:mg_t}
\end{figure}

We characterize the general physical proprieties of the sample by investigating the relation between the mass and the temperature of the gas. 
We consider the error-weighted mean of the temperatures measured in the spectral analysis at radii above 70 kpc. 
The gas mass is calculated by integrating the gas density profile over a spherical volume of radius $R_{500}$ evaluated from the mass profile that we constrain as
discussed in Sec. \ref{sec:methods}. 
We fit the relation
\begin{equation}
 \log \left( \frac{h(z)M_{gas,500}}{M_\odot} \right) = \log N + \tau \log \left( \frac{T}{5 keV} \right) \,
\end{equation}
using the Bayesian regression code LIRA of \citet{sereno16}. 
We obtain $\log N = 13.70 \pm 0.04$ and $\tau=1.98 \pm 0.18$, with an intrinsic scatter $\sigma_{int}=0.134 \pm 0.023$. 
Figure~\ref{fig:mg_t} shows the $M_{gas,500}-T$ relation for the clusters in the sample together with the best-fitting relation, 
compared to the relation found by \citet{arnaud+07} for a sample 10 morphologically relaxed nearby clusters observed with \xmm\ in the temperature range 2--9 keV. 
We find that the two relations are in agreement within the scatter, that in our sample is a factor $\sim4$ higher than the one measured for the 
sample of relaxed local systems in \citet{arnaud+07}. 
The agreement is not improved once only the most ``relaxed'' systems, the ones identified in the upper left quadrant of Fig.~\ref{fig:morphology}, are considered, 
suggesting that more relevant selection biases affect any comparison between our sample and the one in \citet{arnaud+07}.

We also compare the same relation with the gas mass estimated within a radius $R_{2500}$ to the results obtained for the CLASH sample \citep{pos+al12, donahue+14}.  
As shown in Fig.~\ref{fig:mg_t}, our best-fit relation agrees well with the relation derived for the CLASH clusters, with a remarkable agreement on the intrinsic scatter 
($\sigma_{int} = 0.113 \pm 0.008$ for our sample, $0.093 \pm 0.002$ for CLASH).

Overall, we conclude that our sample spanning a wide range both in redshift ($0.41-1.24$) and in the dynamical properties as inferred from proxies based on the X-ray morphology is certainly less homogeneous than the samples of local massive objects, but it is comparable in its physical properties to, e.g., the CLASH systems that were selected just to be X-ray morphologically no disturbed and massive (i.e. very X-ray luminous) at intermediate/high redshifts in a similar manner as we select our targets.

\subsection{On the completeness of the sample}
\label{sec:complete}

\begin{figure}
\begin{tabular}{c}
\includegraphics[width=0.45\textwidth]{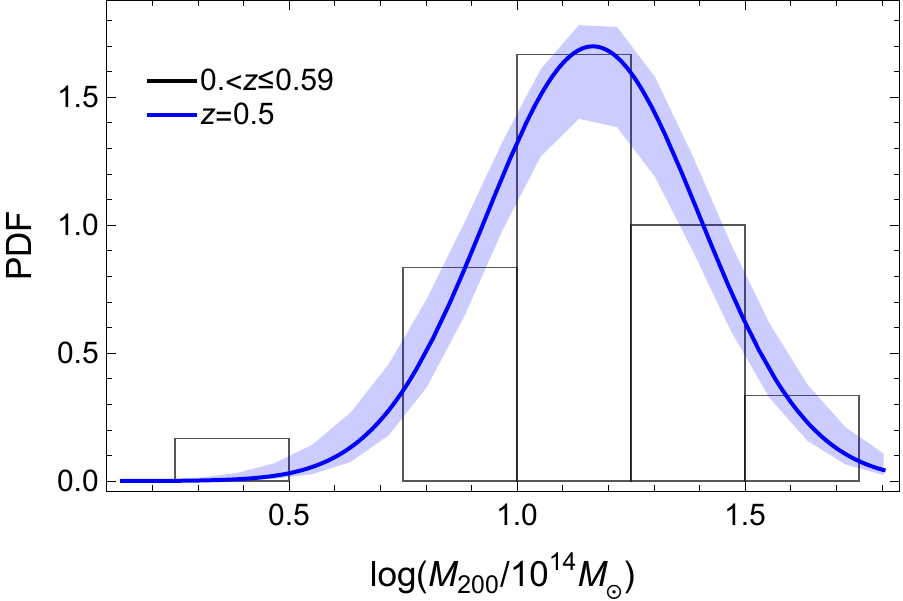} \\
\includegraphics[width=0.45\textwidth]{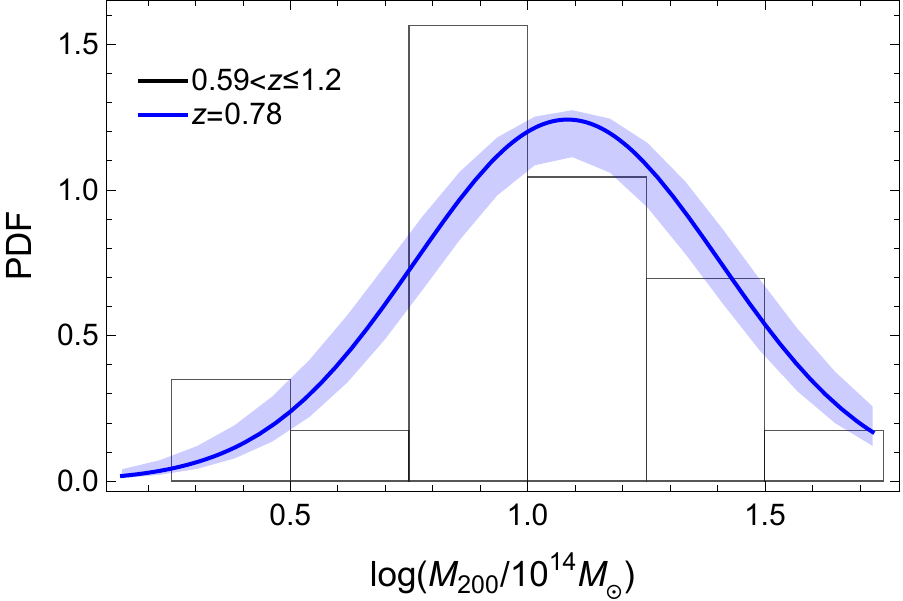} \\
\end{tabular}
\caption{Mass distribution of the selected clusters in two redshift bins. The black histogram bins the observed masses. The blue line is the normal approximation estimated from the regression at the median redshift (see text for details). The shaded blue region encloses the 68.3 per cent probability region around the median relation due to parameter uncertainties. Redshift increases from the top to the bottom panel. The median and the boundaries of the redshift bins are indicated in the legends of the respective panels.}
\label{fig_M200_z_histo}
\end{figure}

\begin{figure}
\resizebox{\hsize}{!}{\includegraphics{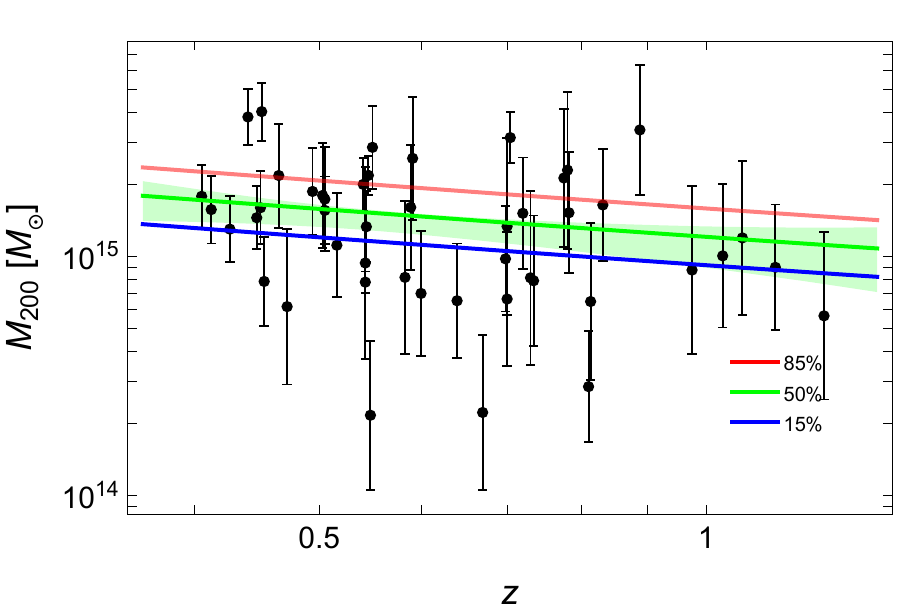}}
\caption{
Completeness functions compared to the distribution of the selected clusters in the $M_{200}-z$ plane (black points). 
The full lines plot the true value of mass (see text and App. A in \citealt{se+et15_comalit_IV} for details) below which a given fraction (from top to bottom: 85, 50, and 15 per cent levels, respectively) of the selected sample is contained. The shaded green region encloses the 68.3 per cent confidence region around the 50 per cent level due to uncertainties on the parameters.}
\label{fig_M200_z_evol}
\end{figure}

Some completeness properties of the selected objects can be studied through the analysis of the mass distribution of the clusters in the sample \citep[see App.~A in][]{se+et15_comalit_IV}. 
As far as the mass distribution is well approximated by a regular and peaked distribution, the completeness of the observed sample can be usually approximated as a complementary error function
\begin{equation}
\label{eq_pmu_1_1}
\chi (\mu) \simeq \frac{1}{2}  \mathrm{erfc} \left( \frac{\mu_\mathrm{\chi}-\mu}{\sqrt{2}\sigma_\mathrm{\chi}}\right),
\end{equation}
where $\mu$ is logarithm of the mass. At first order, $\mu_\mathrm{\chi}$ and $\sigma_\mathrm{\chi}$ can be approximated by the mean and the standard deviation of the observed mass distribution.

We perform the analysis of the completeness together with the $c-M$ relation, as \textsc{LIRA} fits at the same time the scaling parameters and the distribution of the covariate. 
The mass distribution for the observed masses is estimated from the regression output, i.e. the function of the true masses, by smoothing the prediction with a Gaussian whose variance is given by the quadratic sum of the intrinsic scatter of the (logarithmic) mass with respect to the true temperature and the median observational uncertainty.
In Fig.~\ref{fig_M200_z_histo}, we plot the mass distributions in two redshifts bins and we show that the log-normal distribution provides an acceptable approximation to them. 

In Fig.~\ref{fig_M200_z_evol}, we plot the redshift dependence of the selected clusters. The mass completeness limits are fairly constant, differently to survey selected clusters, where the mass limits usually increase with the distance. The selection criteria we imposed on the temperature profile and on the morphological properties effectively selected the very massive high end of the cluster halo function in the investigated redshift range.

Notwithstanding some heterogeneity in the selection criteria, the sample is well behaved, suggesting that the targeted observations of X-ray clusters cover the very luminous end at each redshift. The effective flux threshold decreases with redshift in order to have a number of high redshift objects comparable to the intermediate ones.

\section{Summary and Conclusions}
\label{sec:summary}
We investigate the concentration-mass relation for a sample of $47$ galaxy clusters observed with \chandra\ in the redshift range $0.4<z<1.2$.
We consider the largest sample investigated so far at $z>0.4$ and we provide the first constraint on the $c(M)$ relation at $z>0.7$ from X-ray 
data only. 

We have selected archival exposures of targets with no major mergers and with sufficient X-ray signal to allow to recover properly the hydrostatic mass. 
Using X-ray morphological estimators, we verify that about 1/3 of the sample is not completely relaxed, and that this fraction rises to 0.5 in the objects at $z>0.8$.

As consequence of our selection, the sample is not statistically complete and includes targets that were selected differently for their original observations.
This implies that some unquantifiable bias could be present and could affect the interpretation of the results.
However, we have verified that the sample presents a $M_{\rm gas}-T$ relation that behaves quite similarly to the one estimated locally (see Sect.~\ref{sec:properties}), 
and that, being the selected objects very luminous in the X-ray band, the selection applied is, in practice, on the total mass and 
tends to represent properly the very massive high end of the cluster halo function, in particular at high redshift (see Sect.~\ref{sec:complete}).

We perform a spatial and a spectral analysis for each cluster, and we extract the radial profiles of the gas temperature and density (obtained
from the geometrical deprojection of the surface brightness). We reconstruct the total mass profile by assuming spherical symmetry of the ICM 
and hydrostatic equilibrium between the ICM and the gravitational potential of the cluster, assumed to have a NFW profile described by a 
scale radius $r_\text{s}$ and a concentration $c$. We obtain constraints on $(r_\text{s},c)$ by minimising a merit function in which 
the spectral temperature profile is matched with the temperature predicted from the inversion of the
Hydrostatic Equilibrium Equation that depends only on these parameters.

We are able to determine the temperature profiles up to a median radius of $0.3\,R_{200}$ and the gas density profile up to a median radius of 
$0.5\,R_{200}$. Beyond these limits, and at $R_{200}$ in particular, our estimates are the result of an extrapolation.

Our hydrostatic mass estimates are in very good agreement with the result from weak-lensing analysis available in literature. 
In particular, the c--M relation calculated for the clusters shared with the CLASH sample is fully consistent within the errors.

We estimate a total mass $M_{200}$ in the range (1st and 3rd quartile) $8.1-18.6\times 10^{14} M_\odot$ and a concentration $c_{200}$
between $2.7$ and $5$. The distribution of concentrations is well approximated by a lognormal function in all the mass and redshift 
ranges investigated.

Our data confirm the expected trend of lower concentrations for higher-mass systems and, at a fixed mass range, 
lower concentrations for higher-redshift systems. The fit to the linear function 
$\log c_{200}= A+B\times\log M_{200}/(10^{14}M_\odot)+C\times \log (1+z)\pm \sigma_{\log c_{200}})$
gives: a normalisation $A=1.15 \pm 0.29$, a slope $B=-0.50 \pm 0.20$ which is slightly steeper than the value predicted by 
numerical simulations ($B\sim-0.1$), a redshift evolution $C=0.12 \pm 0.61$ consistent with zero and an intrinsic scatter
on the concentration $\sigma_{\log c_{200}}=0.06 \pm 0.04$.

The predictions from numerical simulations on the estimates of the normalisation $A$ and slope $B$ are in a reasonable agreement with our observational 
constraints at $z>0.4$, once the correlation between them is fully considered (see Fig.~\ref{fig:2DcontoursAB}).
Values from \citet{dm14} are consistent at $1 \sigma$ level. Larger deviations, but still close to the $\sim 2 \sigma$ level of confidence,
are associated to the predictions from \citet{dk15} and \citet{prada+12}, with the latter being the one more in tension with our measurements.

In the redshift range $0.8<z<1.5$, constraints on the $c-M$ relation were also derived in \citet{sc13} for a heterogeneous sample of 31 
massive galaxy clusters with weak and strong lensing signal, obtaining similar results to the ones discussed here, with a slope slightly steeper 
than the theoretical expectation.

It is worth noticing that with this analysis, that represents one of the most precise determination of the hydrostatic mass concentrations in high-z galaxy clusters,
we are characterising the high mass end of the distribution of galaxy clusters even at $z~\sim1$, that is a regime hardly accessible to the present numerical simulations.

A homogeneous sample, and dedicated X-ray follow-up, would improve any statistical evidence presented in our study.
In particular, an extension of this analysis to lower redshifts, still using consistently \chandra\ data, and a careful identification of a sub-sample of the
most relaxed systems would constrain at higher confidence any evolution in the concentration-mass relation for clusters of galaxies, also as function
of their dynamical state.

\section*{ACKNOWLEDGEMENTS}

We thank Benedikt Diemer and the anonymous referee for helpful comments that improved the presentation of the work.
S.E. and M.S. acknowledge the financial contribution from contracts ASI-INAF I/009/10/0 and 
PRIN-INAF 2012 `A unique dataset to address the most compelling open questions about X-Ray Galaxy Clusters'.
M.S. acknowledges also financial contributions from PRIN INAF 2014 `Glittering Kaleidoscopes in the sky: 
the multifaceted nature and role of galaxy clusters'. 
This research has made use of the NASA/IPAC Extragalactic Database (NED) which is operated by the Jet Propulsion Laboratory, 
California Institute of Technology, under contract with the National Aeronautics and Space Administration.

\begin{appendix}  

\section{The observed radial profiles of the gas density and temperature}  
We present here the deprojected density and spectral temperature profiles of all the clusters analysed in this work, as described in 
Sect. \ref{sec:analysis}. 

\begin{figure*}
\centering
\includegraphics[width=0.8\textwidth]{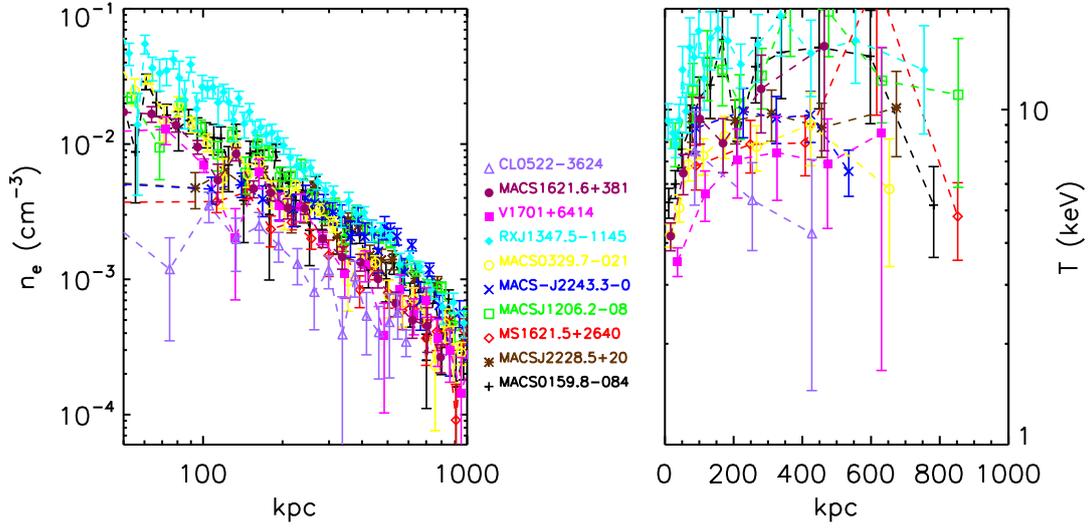}
\caption{Deprojected electron density (left) and spectral temperature (right) profiles for the clusters in the redshift range 0.405-0.472.}
\label{fig:prof0}
\end{figure*}

\begin{figure*}
\centering
\includegraphics[width=0.8\textwidth]{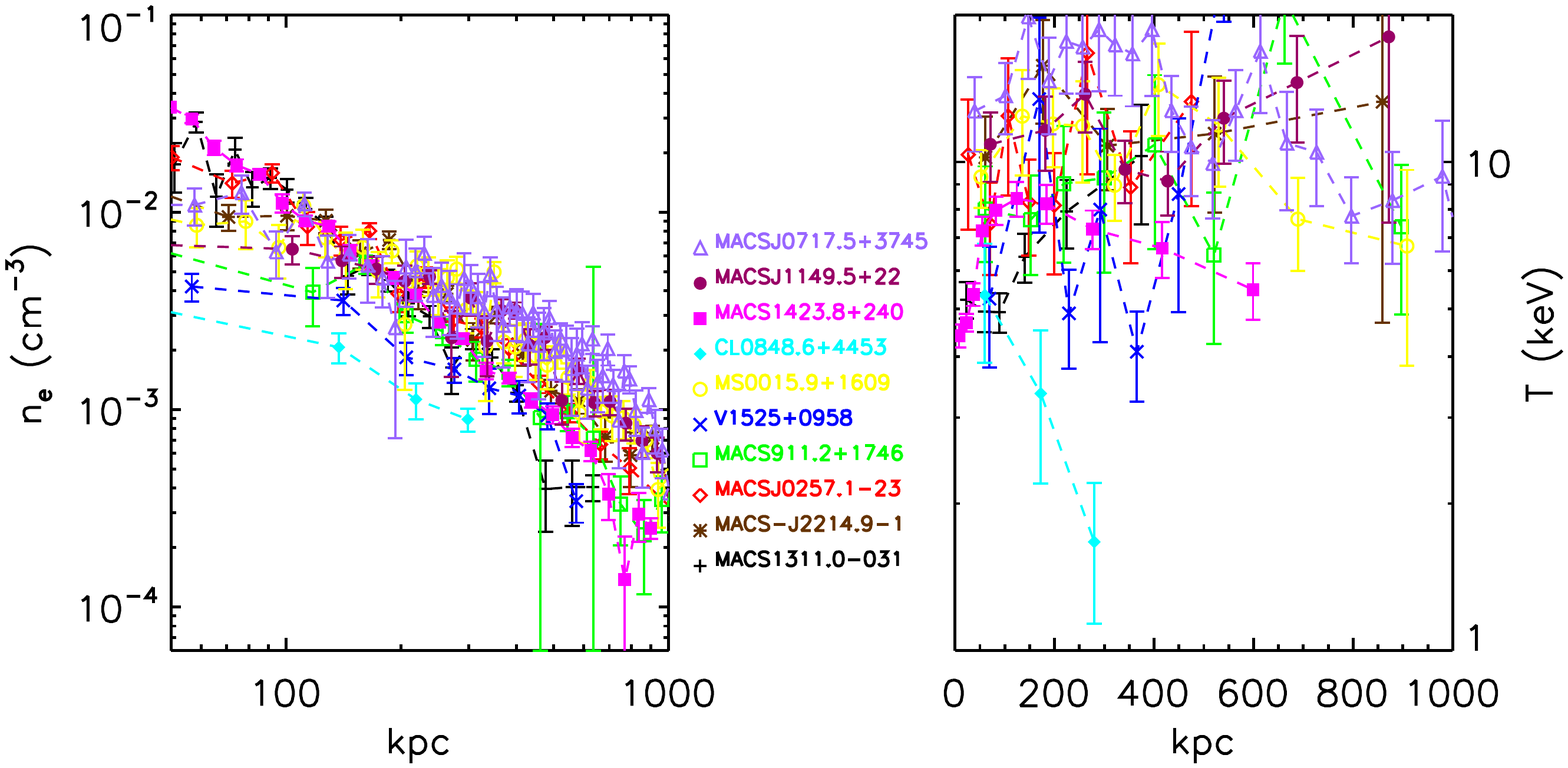}
\caption{The same as in Fig. \ref{fig:prof0} for the clusters in the redshift range 0.494-0.546.}
\label{fig:prof1}
\end{figure*}

\begin{figure*}
\centering
\includegraphics[width=0.8\textwidth]{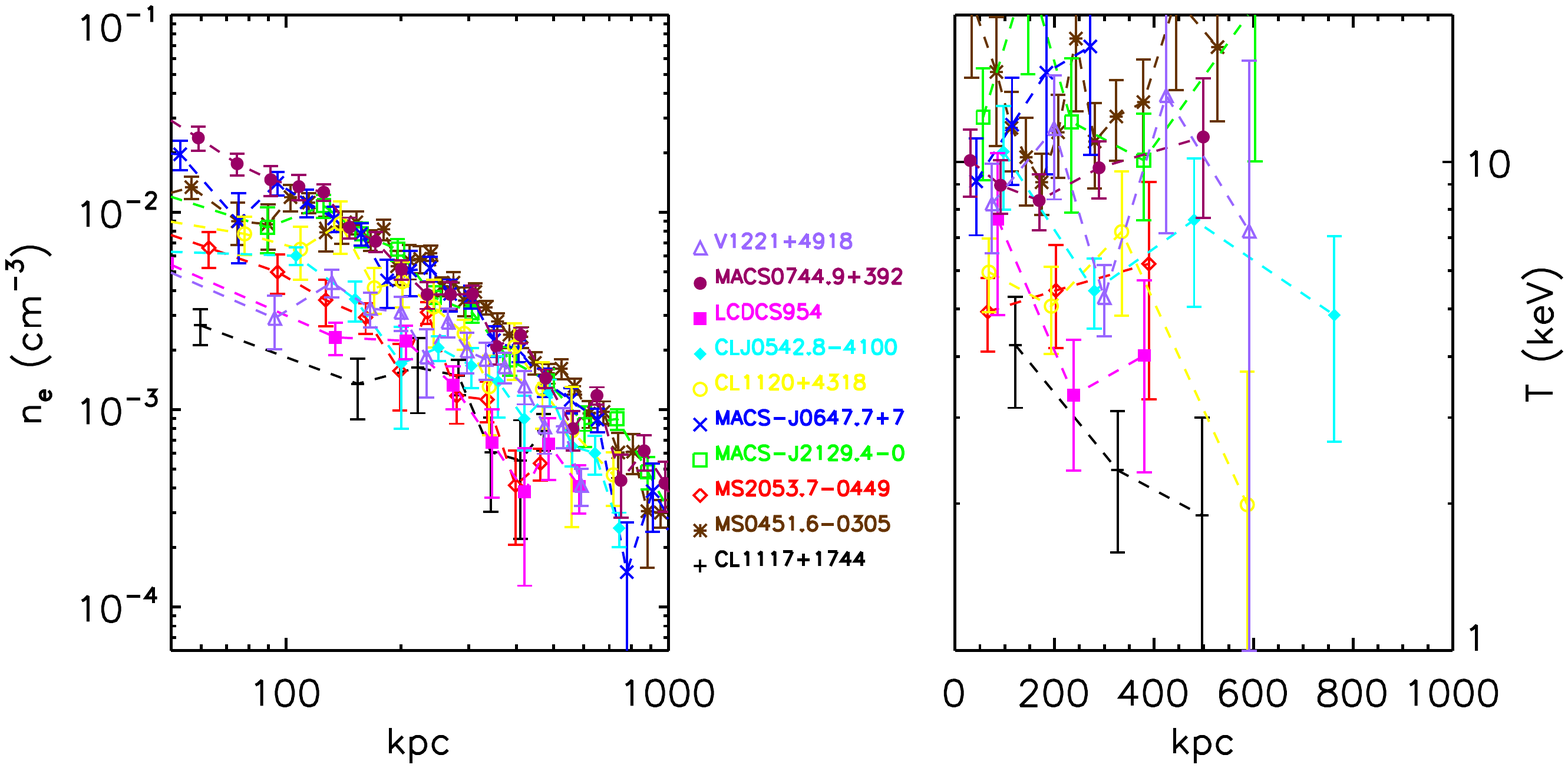}
\caption{The same as in Fig. \ref{fig:prof0} for the clusters in the redshift range 0.548-0.7.}
\label{fig:prof2}
\end{figure*}

\begin{figure*}
\centering
\includegraphics[width=0.8\textwidth]{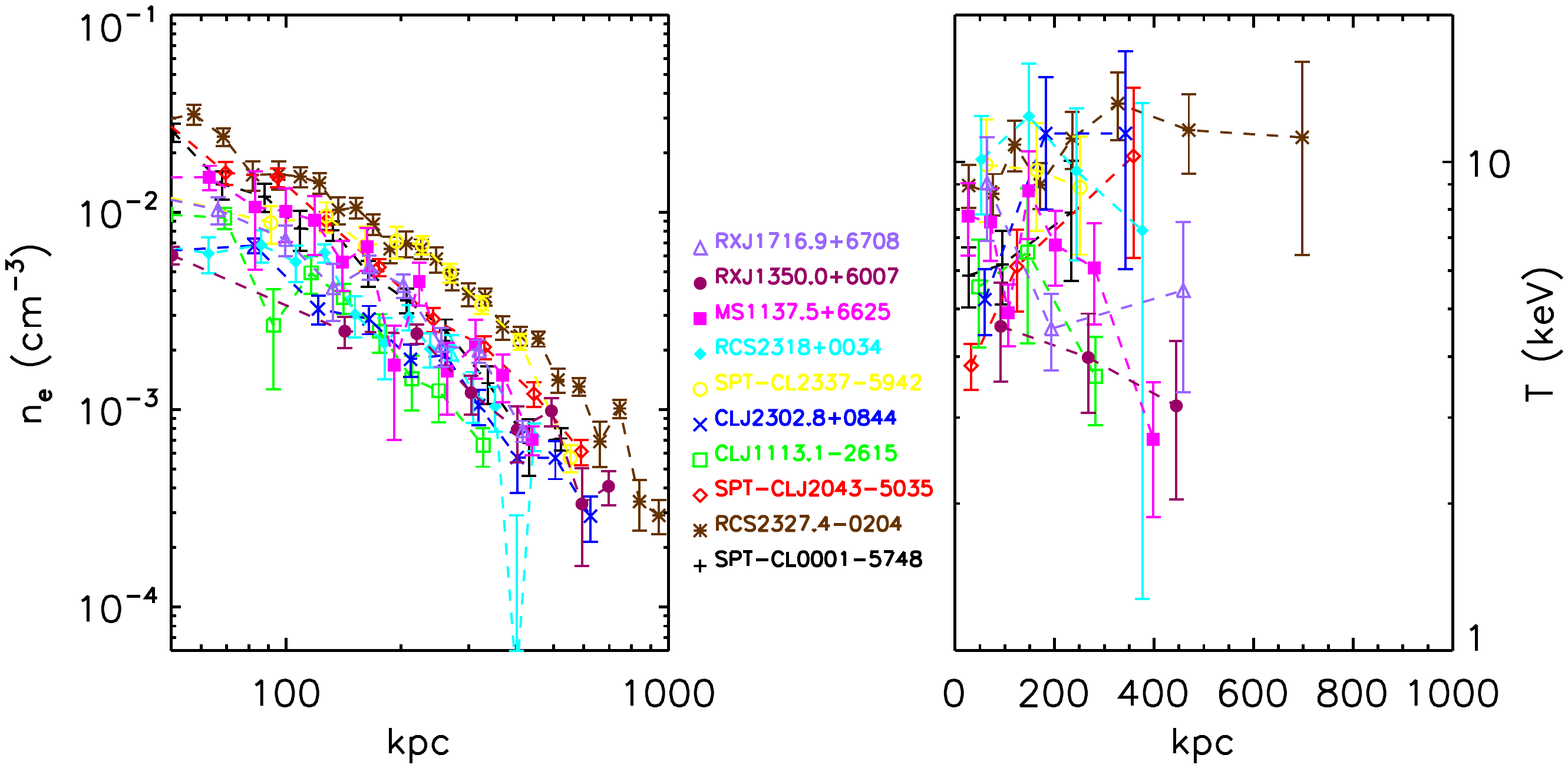}
\caption{The same as in Fig. \ref{fig:prof0} for the clusters in the redshift range 0.7-0.813.}
\label{fig:prof3}
\end{figure*}

\begin{figure*}
\centering
\includegraphics[width=0.8\textwidth]{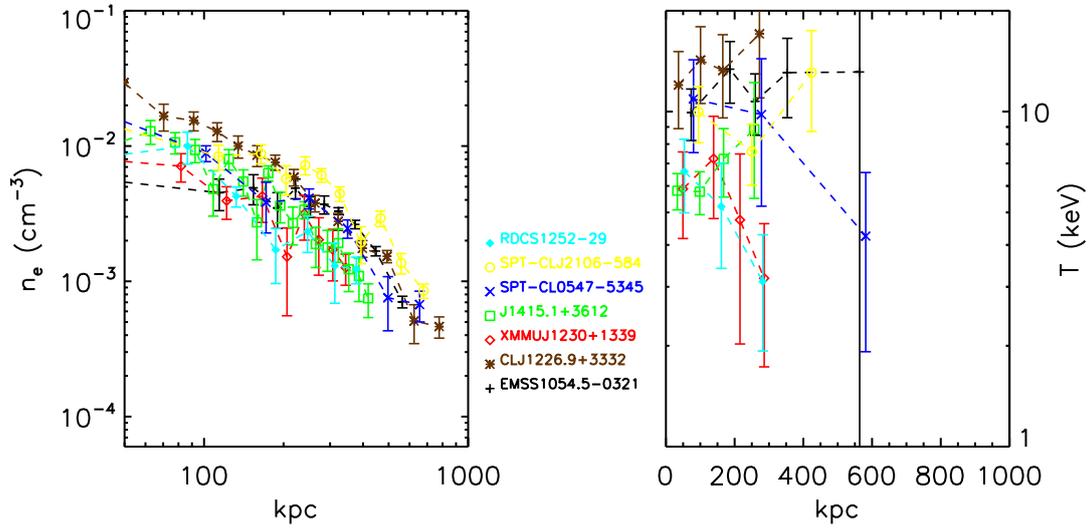}
\caption{The same as in Fig. \ref{fig:prof0} for the clusters in the redshift range 0.831-1.235.}
\label{fig:prof4}
\end{figure*}

\end{appendix}

\end{document}